\documentclass[pre,twocolumn,superscriptaddress,floatfix]{revtex4-1}  %floatfix

\usepackage[paperwidth=210mm,paperheight=297mm,centering,hmargin=2cm,vmargin=2.5cm]{geometry}

\usepackage[pdftex]{graphicx}
\usepackage{amsmath,amsfonts,amssymb}
\usepackage{natbib}
\usepackage{MnSymbol}

\usepackage{hyperref}
\usepackage{xcolor} 
\usepackage{array}
\usepackage{pgfplots}
\usepackage{graphicx}
\usepackage[left,modulo,pagewise]{lineno}
\usepackage{dsfont}

\definecolor{bluemoi}{rgb}{0.25,0.50 ,0.75} 

\hypersetup{
  bookmarksopen=true,
  pdftitle=" ",
  pdfauthor=" ", 
  pdfsubject=" ", 
  pdftoolbar=true, % toolbar hidden
  pdfmenubar=true, %menubar shown
  pdfhighlight=/O, %effect of clicking on a link
  colorlinks=true, %couleurs sur les liens hypertextes
  pdfpagemode=UseNone, %aucun mode de page
  pdfpagelayout=SinglePage, %ouverture en simple page
  pdffitwindow=true, %pages ouvertes entierement dans toute la fenetre
  linkcolor=bluemoi, %couleur des liens hypertextes internes
  citecolor=bluemoi, %couleur des liens pour les citations
  urlcolor=bluemoi %couleur des liens pour les url
}

\makeatletter
\renewcommand{\figurename}{\sf \textbf{Figure}}
\renewcommand{\thefigure}{\arabic{figure}}
\renewcommand{\fnum@figure}{\sf\textbf{\figurename}~\textbf{\thefigure}}
\renewcommand{\tablename}{\sf\textbf{Table}}
\renewcommand{\thetable}{\arabic{table}}
\renewcommand{\fnum@table}{\sf\textbf{\tablename}~\textbf{\thetable}}
\makeatother

\begin{document}

\title{Is spatial information in ICT data reliable?} 

\author{Maxime Lenormand}
\thanks{Corresponding author: maxime.lenormand@irstea.fr}
\affiliation{Irstea, UMR TETIS, 500 rue Fran\c cois Breton, FR-34093 Montpellier, France}

\author{Thomas Louail}
\affiliation{Instituto de F\'isica Interdisciplinar y Sistemas Complejos IFISC (CSIC-UIB), Campus UIB, ES-07122 Palma de Mallorca, Spain}
\affiliation{CNRS, UMR 8504 G\'eographie-Cit\'es, 13 rue du four, FR-75006 Paris, France}

\author{Marc Barthelemy}
\affiliation{Institut de Physique Th\'{e}orique, CEA-CNRS (URA 2306), FR-91191, Gif-sur-Yvette, France}
\affiliation{CAMS, EHESS-CNRS (UMR 8557), 190-198 avenue de France, FR-75013 Paris, France}

\author{Jos\'e J. Ramasco}
\affiliation{Instituto de F\'isica Interdisciplinar y Sistemas Complejos IFISC (CSIC-UIB), Campus UIB, ES-07122 Palma de Mallorca, Spain}

\begin{abstract} 
  An increasing number of human activities are studied using data
  produced by individuals' ICT devices. In particular, when ICT
  data contain spatial information, they represent an invaluable
  source for analyzing urban dynamics. However, there have been
  relatively few contributions investigating the robustness of
  this type of results against fluctuations of data
  characteristics. Here, we present a stability analysis of
  higher-level information extracted from mobile phone data
  passively produced during an entire year by 9 million
  individuals in Senegal. We focus on two information-retrieval
  tasks: (a) the identification of land use in the region of
  Dakar from the temporal rhythms of the communication activity;
  (b) the identification of home and work locations of anonymized
  individuals, which enable to construct Origin-Destination (OD)
  matrices of commuting flows. Our analysis reveal that the
  uncertainty of results highly depends on the sample size, the
  scale and the period of the year at which the data were
  gathered. Nevertheless, the spatial distributions of land use
  computed for different samples are remarkably robust: on
  average, we observe more than 75\% of shared surface area
  between the different spatial partitions when considering
  activity of at least 100,000 users whatever the scale. The OD
  matrix is less stable and depends on the scale with a share of
  at least 75\% of commuters in common when considering all types
  of flows constructed from the home-work locations of 100,000
  users. For both tasks, better results can be obtained at larger
  levels of aggregation or by considering more users. These
  results confirm that ICT data are very useful sources for the
  spatial analysis of urban systems, but that their reliability
  should in general be tested more thoroughly.
\end{abstract}

\maketitle

\begin{figure*}
  \centering 
  \includegraphics[width=\linewidth]{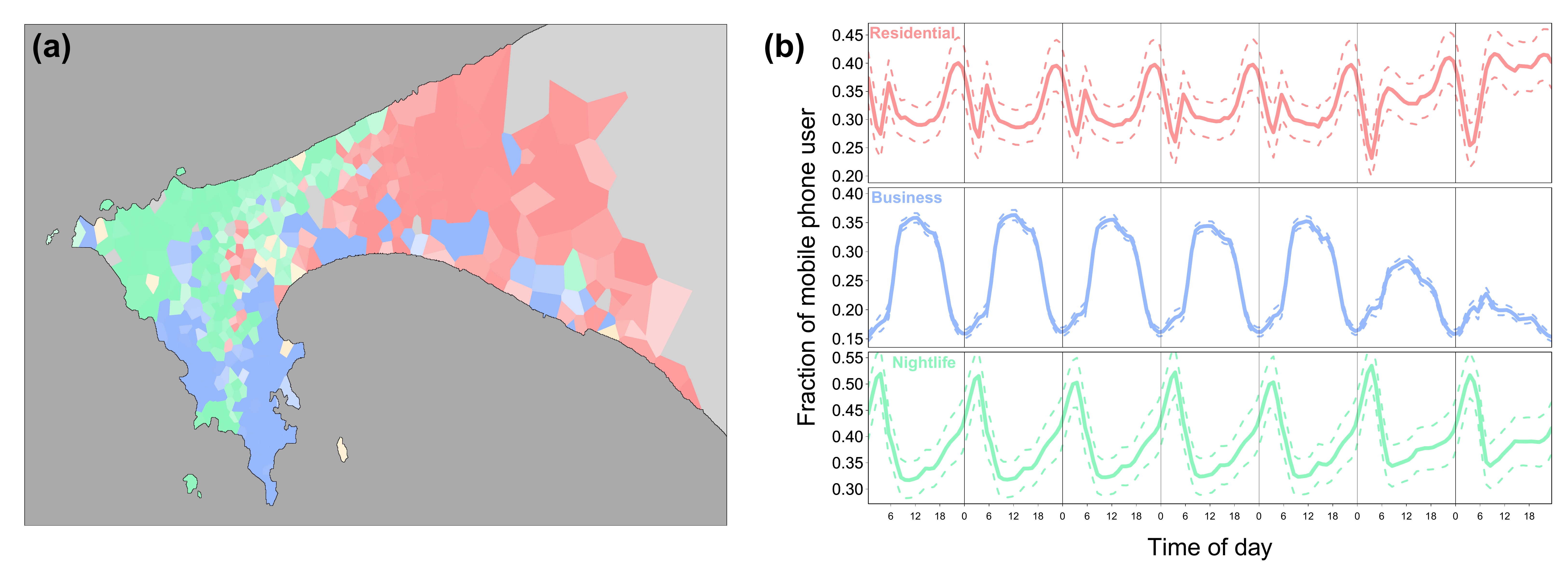}
  \caption{\textbf{Spatio-temporal uncertainty propagation while
      inferring land use from mobile phone activity.} (a) Map of
    the region of Dakar displaying the three clusters according
    to their land use. Colors vary from white to the most
    recurrent cluster identified in the random samples. The color
    saturation depends on the number of times the zone was
    classified as the most recurrent cluster. The color code is
    red for \emph{Residential}, blue for \emph{Business}, green
    for \emph{Nightlife} and orange for other types of
    land use. (b) Temporal patterns associated with the three
    clusters. The solid lines represent the average temporal
    profile computed over the random extractions, while the
    dashed lines represent one standard deviation. These results
    were obtained at the Voronoi scale for 50 independent random
    extractions of 150,000 users mobile phone activity during one
    week. \label{Fig1} }
\end{figure*}

\section*{Introduction}

Massive amounts of geolocalized data are passively and
continuously produced by individuals when they use their mobile devices: smart phones, credit cards, GPSs, RFIDs or remote sensing devices. This deluge of digital footprints is growing at an extremely fast pace and represents an unprecedented opportunity for researchers, to address quantitatively challenging questions, in the hope of unveiling new insights on the dynamics of human societies. Many fields are concerned by the development of new techniques to handle these vast datasets, and range from applied mathematics, physics, to computer science, with plenty of applications to a variety of disciplines such as medicine, public health and social sciences for example.

Although data resulting from the use of information and
communications technologies (ICT) have the advantage of large
samples sizes (millions of observations), and high
spatio-temporal resolution, they also raise new challenging
issues. Some are technical and related to the storage, management
and processing of these data \cite{Kaisler2013}, while others are
methodological, such as the statistical validity of analysis
performed on such data. For example, in the case of mobile phone
data, researchers have often no control and limited information
regarding the data collection process, which obviously deserves
other purposes than scientific research. Various hidden biases
can affect these data used to study the spatial behavior of
anonymized individuals, and consequently observing the world
through the lenses of ICT data may therefore lead to possible
distortions and erroneous conclusions \cite{Lewis2015}. It is
thus crucial to perform statistical tests and to develop methods
in order to assess the robustness of the results obtained with
ICT data. In the research community that studies human mobility
in urban contexts
\cite{Ratti2006,Louail2014,Calabrese2015,Louail2015}, efforts in
this sense have been made in recent years, notably by
cross-checking results \cite{Lenormand2014} obtained with ICT
data and with more traditional data sources
\cite{Schneider2013,Tizzoni2014,Lenormand2014,Deville2014,Alexander2015,Toole2015,Jiang2016}. These
comparisons cover different topics, such as the analysis of
daily mobility motifs \cite{Schneider2013}, the distribution of
population at different scales \cite{Lenormand2014,Deville2014},
the estimation of commuting flows
\cite{Tizzoni2014,Lenormand2014,Alexander2015,Jiang2016}, and the
identification of land uses
\cite{Lenormand2014,Toole2015}. However, the robustness of
results to sample selection, scale or sample size has, up to our
knowledge, never been studied so far.

In the following, we present two examples of such uncertainty
analysis on higher-level spatial information extracted from
mobile phone metadata, which were produced in Senegal in 2013
\cite{Montjoye2014}. We concentrate on two information-retrieval
tasks: first, we evaluate the uncertainty when inferring land use
from the rhythms of human communication
\cite{Soto2011,Frias2012,Toole2012,Pei2013,Lenormand2015};
second, we quantify the uncertainty when identifying individuals'
most visited locations
\cite{Ahas2010,Isaacman2011,Lenormand2014,Toole2015}. We conclude
by mentioning possible future steps in order to assess more
clearly the relevance of various ICT data sources for studying
a variety of urban dynamics.

\section*{Study area and data description}

We focus here on the region of Dakar, Senegal. The mobile phone
data consists in call detail records (CDR) of phone calls and
short messages exchanged by more than 9 million of anonymized
Orange's customers. They were collected in Senegal in 2013, and
were released to research teams in the framework of the 2014
Orange Data for Development challenge \cite{Montjoye2014}. We use
in this study the second dataset (SET2) that was made available
by Orange, and which contains fine-grained location data on a
rolling 2-week basis at the individual level. For each of the 25
two-weeks periods, a sample of about 300,000 mobile phone users
were randomly selected at the country scale. Whenever one of
these individuals uses his/her mobile phone during the two-week
period, the time and his/her position (at the level of serving
cell tower) are recorded. These information can be used to study
human activity and mobility patterns in the region of Dakar, that
is here divided into 457 zones. This spatial partition is the
Voronoi tessellation constructed from the location of phone
antennas in the city, chosen as nodes. Each Voronoi cell thus
approximates the activity zone served by the antenna located at
its center (see Figure S1a).

\begin{figure*}
  \centering 
  \includegraphics[width=\linewidth]{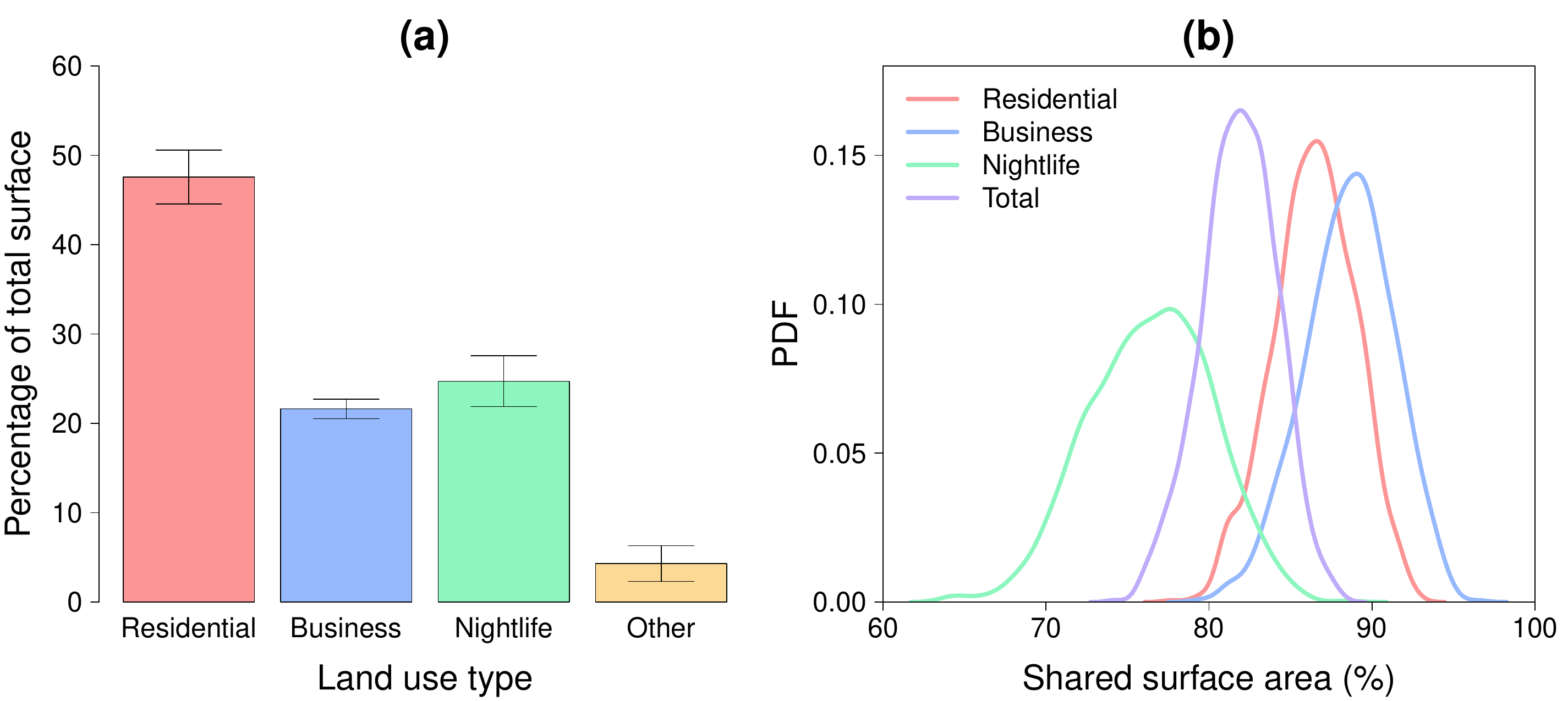}
  \caption{\label{Fig2} \textbf{Uncertainty when inferring
      land use from mobile phone activity.} (a) Area covered by
    the different land use types, expressed as a percentage of
    the total surface. The values have been averaged over 50
    random extractions, and the error bars represent one standard
    deviation. (b) Probability density function of the shared
    surface area between each pair of spatial distributions
    according to the type of land use. These results were
    obtained at the scale of the mobile phone towers (Voronoi
    tesselation), and they are based on the comparison of 50
    $\times$ 50 independent samples with 150,000 signals.}
\end{figure*}

\section*{Inferring land use from mobile phone activity}

\subsection*{Functional network of the city}

Geolocalized ICT data have been widely used to infer land use
from human activity
\cite{Soto2011,Frias2012,Toole2012,Pei2013,Lenormand2015}. The
basic idea is to divide the region of interest into zones,
extract a temporal activity signal for each of these
zones, and then cluster together zones that display similar
signals. Each of the resulting clusters then corresponds to a
certain type of activity (\emph{Residential}, \emph{Commercial},
\dots). We used here the functional approach proposed in
\cite{Lenormand2015}. This method takes as input, for each zone,
a signal composed of 168 points (24h $\times$ 7 days), each value
corresponding to the number of users located in this zone, at
this hour of the day, this day of the week. These signals are
then normalized by the total hourly activity, in order to
subtract trends introduced by circadian rhythms. A Pearson
correlation matrix between zones is then computed. Two zones
whose activity rhythms are strongly correlated in time will have
a high positive correlation value. This similarity matrix can be
represented by an undirected weighted network, which is then
clustered using the \emph{Infomap} community detection algorithm
\cite{Rosvall2008}. This method has the advantage to be
non-parametric (the number of clusters is not fixed \textit{a
  priori}).

\subsection*{Signal extraction and sampling strategy}

In order to apply this functional approach to the region of
Dakar, we first need to define a method for sampling and
aggregating spatially the users' mobile phone activity as
extracted from the raw data. In this dataset, the mobile phone
activity in Senegal during the year 2013 has been divided into 25
two-weeks periods that we separate into 50 time windows of one
week. For each week, we build the users' temporal mobile phone
activity by relying on the following criteria: each individual
counts only once per hour. If a user is detected in $k$ different
zones within a given 1-hour time period, each registered position
will count as ($1/k$) `units of activity' for each of these $k$
zones. % Finally, the mobile phone activity of one user during one
% week consists in a temporal signal composed of 168 points
% associated with a fraction of activity and a location.
It is important to note that only signals containing activity in
the region of Dakar have been considered. At the end of the
process, we obtain the temporal signal of about 160,000 users per
week, a value that is quite stable over the 50 weeks. These
temporal signals can then be averaged into a temporal signal of
activity for each zone allowing us to identify different land use
type in the region of Dakar by applying the above method. Note
that since in the original dataset the year is divided into
two-weeks periods and that a user could appear in two or more
two-weeks periods, several signals of a same user (during
different weeks) can be observed among the 50 weeks of activity.

We also need to define a sampling strategy for assessing the
robustness of land use identification with respect to sample
selection. In order to analyze and compare spatial distributions
of land use obtained with different samples of individual
temporal signals, we needed to ensure that these samples are
independent (i.e. no signals in common) and also that they are
evenly distributed across the entire year. For example, if we
want to compare two spatial distributions of land use based on
two independent samples composed of 150,000 signals
(i.e. individuals) each, we will draw at random two independent
samples of 3,000 signals (individuals) for each week. We will
then spatially and temporally aggregate the signals over the 50
weeks in order to obtain two independent aggregate signals, each
composed of 3,000 $\times$ 50 distinct individual signals. We
then apply the functional approach described above for extracting
the spatial partitions of land use, and for investigating the
uncertainty of these partitions to individuals' sample
selection. The influence of the sample size or the scale on the
uncertainty can also be investigated by varying the number of
signals for each random extraction, and/or by spatially
aggregating them over spatial grids made of regular square of
varying (parameterized) sizes.

\subsection*{Spatial propagation of uncertainty}

As a first step, we applied the functional approach on 50
independent random extractions of 150,000 individual signals
across the entire year, at the scale of the Voronoi cells (mobile
phone towers locations), by using the sampling strategy described
in the previous section. Three clusters of zones emerged
systematically, covering on average 95\% of the total
surface. The remaining 5\% correspond to other clusters with no
clear patterns, probably associated with some local punctual
events. We show on Figure \ref{Fig1}b the average temporal
profiles along with the variability around this average, for each
of these three clusters. Each of the clusters can be roughly
associated to a typical rhythm of human activity, and
consequently to a characteristic land use:
\begin{itemize}
\item{} A \emph{Residential} activity profile, corresponding to a
  high probability of mobile phone use during early mornings,
  evenings and week end days.
\item{} A \emph{Business} cluster, displaying a significantly
  higher activity from 9am to 5-6pm during weekdays.
\item{} A \emph{Nightlife} activity profile, characterized by a
  high activity during night hours (1am-4am).
\end{itemize}

The \emph{Nightlife} cluster (in green) covers the area of the
international airport, and also the neighborhood of \emph{`La
  Pointe des Almadies'}, where mainly wealthy people live and
where are located most of the rich nightclubs. The \emph{Business}
cluster covers Dakar's central business district (\emph{`Le
  Plateau'}), where one finds companies' headquarters, and where
the port is also located. Finally, the \emph{Residential} cluster covers
the rapidly growing parts of the Dakar peninsula, which profits
from the highway construction. It is worth noting that the
different land use types identified in this study are consistent
with the ones obtained with another mobile phone dataset in Spain
\cite{Lenormand2015}, except that in the case of Dakar, the
method is not able to distinguish between industrial (or
logistic) and leisure/nightlife activities (see
\cite{Lenormand2015} for more details).

As can be observed in Figure \ref{Fig2}a, the area covered by the
different types of land use is quite stable over the 50 samples,
with the \emph{Residential} land use type representing on average
about 50\% of the total surface, while we observe about 20\% and
25\% for the \emph{Business} and \emph{Nightlife} clusters,
respectively. Nevertheless, the stability of the proportion does
not imply that they follow the same spatial distribution from one
sample to another. In order to test the stability, we computed
the proportion of surface area shared by two spatial
distributions $p_l$ and $p_l'$ of a given type $l$, as obtained
with two different samples. The expression for this quantity is
\begin{equation}
   S=2\frac{A_{p_l \cap p_l'}}{A_{p_l}+A_{p_l'}} ,
	 \label{SSA}
\end{equation}
where $A_{p_l}$ denotes the surface area of spatial distribution
$p_l$. Note that in our case $A_{p_l} \simeq A_{p_l'}$ (Figure
\ref{Fig2}a). Similarly, we can define the total surface area
shared by two spatial partitions $p$ and $p'$ (with the same
number and type of land use) of the region of interest,
\begin{equation}
   S^*=\frac{\sum_{l} A_{p_l \cap p_l'}}{\sum_{l} A_{p_l}} .
	 \label{SSA2}
\end{equation}
The results are displayed in Figure \ref{Fig2}b. The similarities
between the 50 different spatial partitions is globally high,
with on average 80\% of shared surface area. The agreement is
larger for the \emph{Residential} and \emph{Business} clusters
with an average shared surface area around 90\%, against 75\% for
the \emph{Nightlife} land use type. This is probably due to the
more episodic character of the nightlife activity, implying a
smaller statistical reliability of the results. A map of the
region of Dakar displaying the uncertainty associated with the
land use identification is shown in Figure \ref{Fig1}a. The
colors represent the different land use types, and each zone has
been assigned its recurrent cluster type over the 50 land use
identifications. The color saturation is then related to the
uncertainty, quantified by the number of times the zone was
classified as a given recurrent cluster: the color is darker if
the uncertainty is low, paler otherwise. Most of the zones have
been assigned to the same clusters more than 80\% of the time.

\subsection*{Influence of scale and sample size on the
  uncertainty}

\begin{figure}[!h]
  \centering 
  \includegraphics[width=7.5cm]{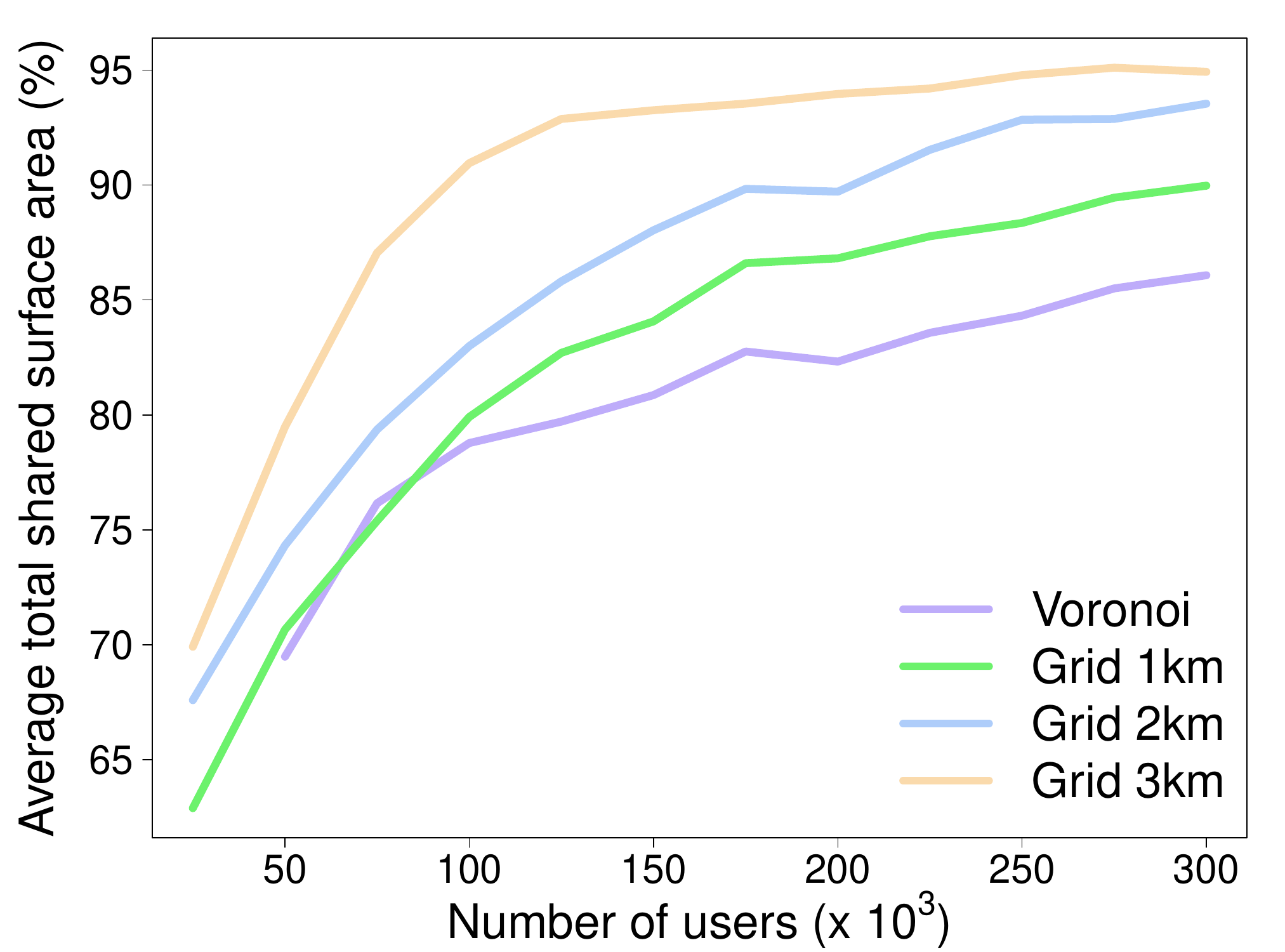}
  \caption{\textbf{Influence of spatial scale and sample size.}
    Total shared surface area $S^*$ between land use partitions as a
    function of the sample size and for different spatial scales
    (Voronoi cells, and square grid cells of resp. 1, 2 and 3 km
    side length). In almost every case, the values displayed have
    been averaged over 100 independent comparisons. See Figure
    S3a for more details, while the full distributions are shown
    on Figure S2 in Appendix. \label{Fig3}}
\end{figure}

The identification of land use from mobile phone activity seems
to be quite robust to the sampling of individuals. So far we
considered communication data of 150,000 users, available at
their maximal spatial resolution, that is the locations of the
mobile phone carrier's antennas. But what about the influence of spatial scale and sample size on this uncertainty? In order to answer this question, we applied the
same functional approach, but this time by (a) varying the number
of individual signals of each random extraction (from 25,000 to
300,000, by step of 25,000) and (b) aggregating them spatially
over spatial square grids with cells of side length equal to 1, 2
or 3 km (Figure S\ref{Fig1}) (the spatial aggregation is based on
the area of the intersection between the Voronoi and the grid
cell) . We performed the comparison between 100 pairs of
independent samples for different population sizes, and spatial
resolutions/grid sizes. The results are shown in Figure
\ref{Fig3}.

Interestingly, the \emph{Nightlife} cluster does not
systematically emerge, especially when considering small sample
sizes at the `Voronoi scale'. In these cases, only two clusters are detected, the \emph{Nightlife} activity cluster is mixed with \emph{Residential} and \emph{Business}
activity clusters. As it can be observed
in Figure S3 in Appendix, at least 100,000 individuals signals
need to be aggregated to detect three clusters in more than 90\%
of the random sample extractions. This quantity falls down at
60\% for sample population sizes of 75,000, about 15\% for 50,000
and 0 for 25,000. We note that it never happened when the
signals have been spatially aggregated. Considering only the partitions for which three
clusters were detected, we observe that the percentage of share
surface area increases when the data are spatially
coarse-grained, and also with the number of individuals taken
into account, revealing the existence of a typical scale. The
order of this scale seems to be here of the order 100,000, above
which we obtain more than 75\% similarity between the land use
spatial partitions. Coarse-graining the spatial resolution by
projecting the data on grids of larger cells plays also an
important role, allowing us to reduce the uncertainty when small
samples are considered. As expected, the variability of the
uncertainty tends to increase with the grid cell side length (see
Figure S2 in Appendix).

\subsection*{Temporal variations}

\begin{figure}[!h]
  \centering 
  \includegraphics[width=7.5cm]{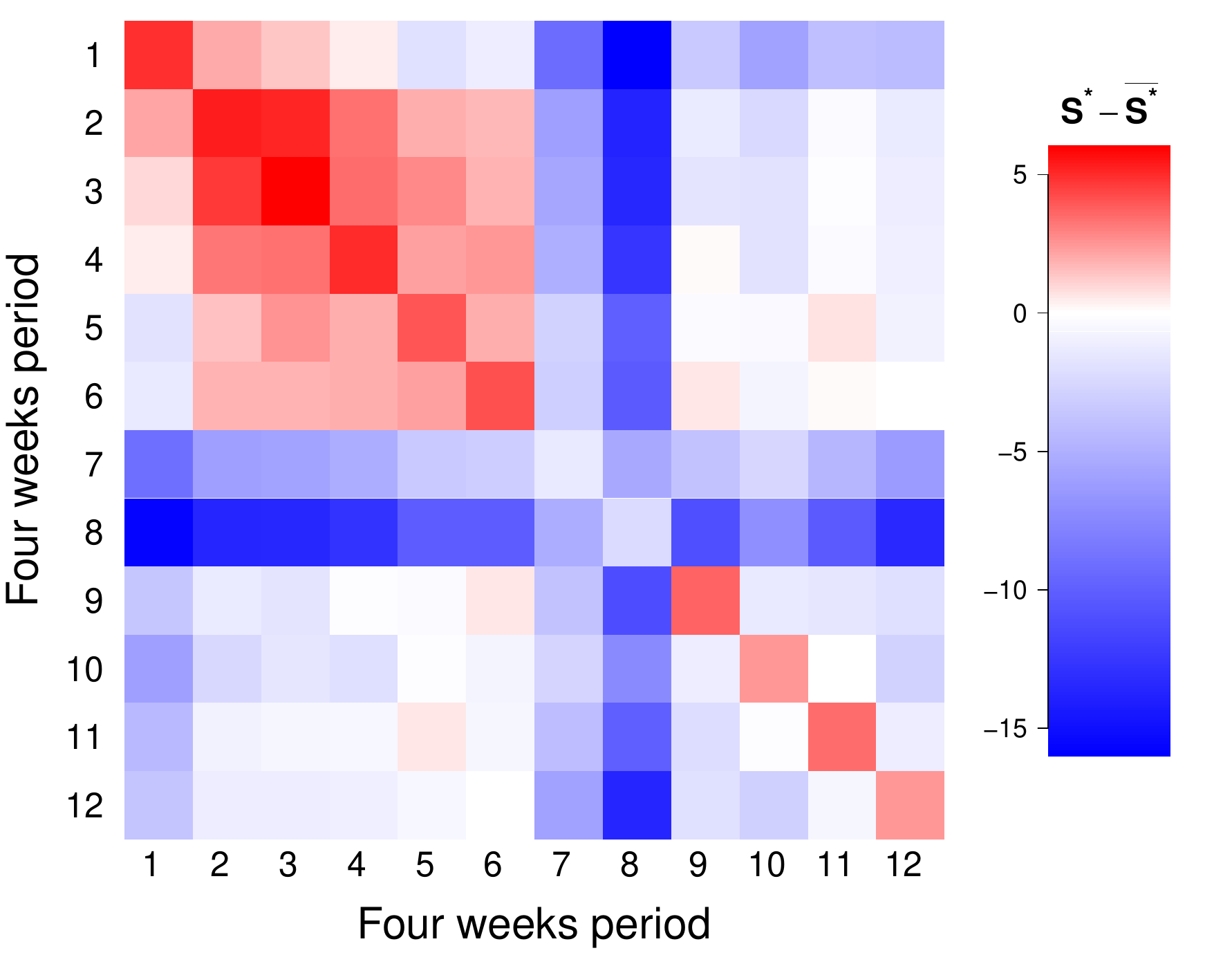}
  \caption{\textbf{Temporal variations.} Standard shared surface
    area between spatial partitions of land use extracted for
    different time windows (four weeks periods, from the first
    week of January to the last week of November). The
    standardization is performed by subtracting the average
    shared surface area obtained by comparing 50 independent
    samples based on 150,000 signals drawn at random across the
    entire year (Figure \ref{Fig2}b). The values have been
    averaged over 100 independent comparisons based on 150,000
    signals obtained at the Voronoi scale. \label{Fig4}}
\end{figure}

We considered so far samples uniformly distributed across the
entire year, but comparisons of samples extracted from different
time windows (i.e. communication activity recorded at different
periods of the year) can also be performed in order to identify
potential temporal variations. We therefore considered 12
consecutive time windows of four weeks (from the first week of
January to the last week of November). For each possible pair of
time-windows, we performed comparisons between 100 pairs of
independent samples, each constituted of 150,000 signals (at the
Voronoi scale). Figure \ref{Fig4} shows the average shared
surface area, standardized by the average shared surface area
across the entire year.

We observe that, in most cases, the shared surface area
is close to the one obtained on average for the entire year. As
expected it is globally lower for the comparisons inside the same
time window, and higher for the comparisons of partitions
extracted for distinct time windows. It is also worth noting that
similarity between time periods decreases with the time elapsed
between them, and that the land use patterns identified in the
first half of the year seems to be more similar than the ones
observed in the second half of the year. It is however not clear
whether these changes are due to changes in the city structure
itself, or to seasonal variations. The most unexpected result is
the change of behavior observed in summer during the weeks 7 and
8 (from the end of June to the middle of August) showing a
similarity always lower than the average, -15 points in the worth
case. This can be explained by the change of activity generally
observe in summer. This information should be nevertheless taken
into account when analyzing this type of data.

\begin{figure*}[!ht]
  \centering 
  \includegraphics[width=\linewidth]{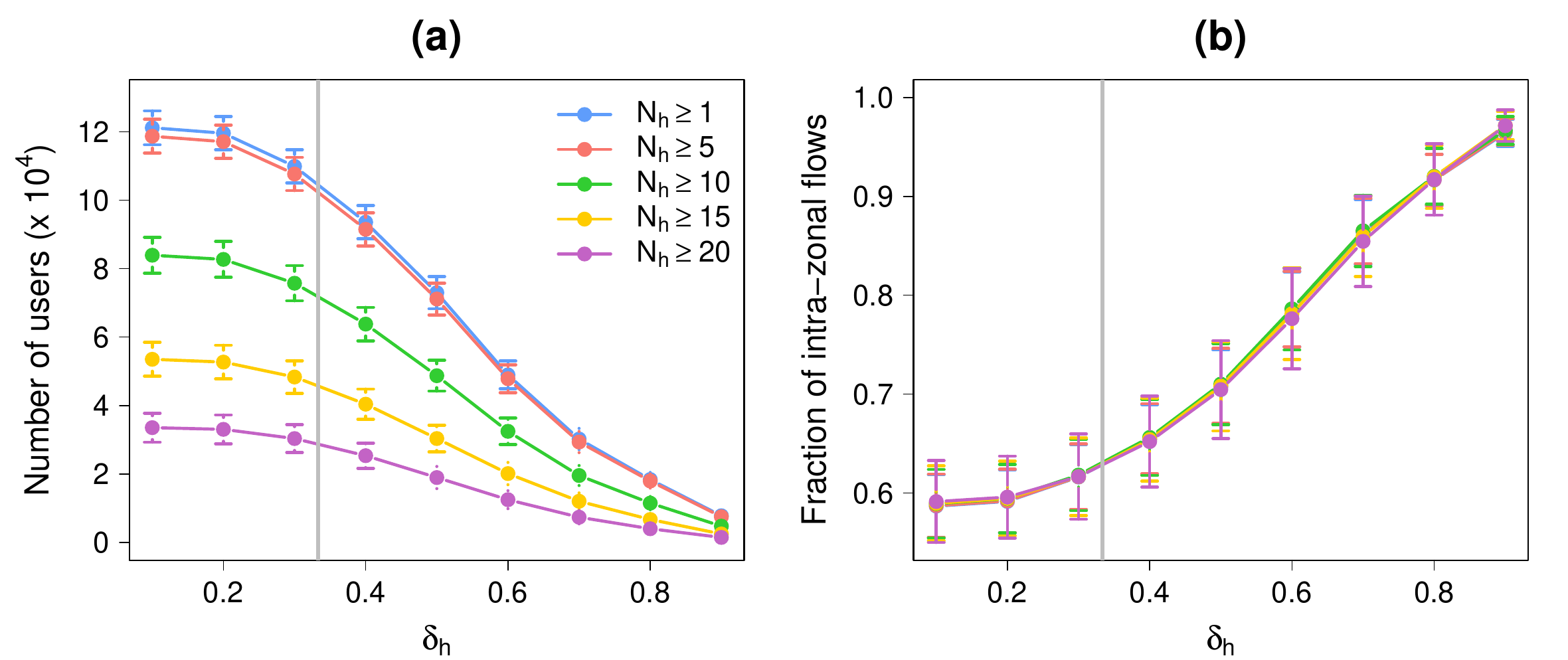}
  \caption{\textbf{Influence of the parameters.} Number of
    reliable users (a) and fraction of users living and working
    in the same zone (b) as a function of $\delta_h$ and for
    different values of $N_h$ ($N_d=4$). Only people living and
    working in the region of Dakar have been considered
    here. These results have been averaged over the 25 two-weeks
    periods, and error bars represent one standard deviation. The
    vertical bars indicate the value $\delta=1/3$. Similar plots
    for different value of $N_d$ ranging from 1 to 8 are shown in
    Figures S4 and S5 in the Appendix. \label{Fig5}}
\end{figure*}

\section*{Identifying home and work locations from mobile phone activity}

\subsection*{Extracting individuals' most visited locations}

Geolocalized ICT data are also widely used to identify the most
visited locations of an individual, allowing to extract the origin-destination (OD) matrices of commuting flows, a fundamental object in mobility studies. A very simple
heuristic used in most methods is the following: the most visited
place of an individual in the late afternoon/evening and in the
early morning is used as a proxy for his/her place of residence,
while the most visited location during working hours is a proxy
for his/her workplace (or main activity place). This simple
assumption has allowed the accurate determination of mobility
flows at intermediate geographical scales for a variety of cities
worldwide (see for example
\cite{Tizzoni2014,Lenormand2014,Alexander2015,Toole2015}). However,
the robustness of the results obtained with such a simple
heuristic, with respect to sample selection, has never been
investigated.

For each of the 25 two-weeks periods and for each user, we apply
the following procedure to extract the home and work locations:
\begin{itemize}
\item First, the hours of activity are divided into two groups,
  daytime hours (between 8am and 5pm included) and nighttime
  hours (between 7pm and 7am included). Only days of the week
  from Monday to Thursday are taken into account, that is to say
  8 days in total over two weeks.
\item We then apply a \textbf{first filter} by considering only
  users who were active at least $N_d$ days (out of 8) during
  daytime \textbf{and} $N_d$ days during night time. 
\item A \textbf{second filter} is then applied to keep only users
  who have been `active', in total during the two weeks, at least $N_h$ 1-hour periods during
  daytime, \textbf{and} $N_h$ 1-hour periods during nighttime.
\item For each hour of activity, the most frequently visited zone during this
  hour is identified (based on his/her geolocalized mobile phone
  activity).	
\item For both groups of hours (daytime and nighttime), we
  identify the zone in which the user has been localized the
  highest number of hours.
\item A \textbf{third and last filter} is also implemented to
  select only users whose fraction of hours spent at `home' and
  `work' are larger than a fraction $\delta_h$ of the total
  number of locations visited during nighttime and daytime,
  respectively.
\item Finally, only individuals living and working in the region
  of Dakar are considered as reliable users.
\end{itemize}

The last filter allow us to adjust the degree of confidence in
the identification of the main nighttime (`Home') and daytime
activity (`Work') locations. A low value of $\delta_h$ allows us
to maximize the number of reliable users (Figure \ref{Fig5}a),
with the risk of including places of main activity where the user
has spent very little time. On the contrary, a high value of
$\delta_h$ will make us keep only users moving very little during
nighttime and/or daytime, and who therefore may not be
representative of the variety of mobility patterns at a small
scale. As it can be observed in Figure \ref{Fig5}b, these users
are generally people living and working in the same area. Based
on these considerations, we chose to fix $\delta_h$ to 1/3 which seems to be a good interplay, allowing us to remove users
exhibiting irregular mobility patterns during the time period,
while preserving the commuting network structure in the Dakar
region.

The behavior of $\delta_h$ on the number of reliable users and
the fraction of intra-zonal flows (users living and working in
the same area) is independent of $N_h$ and $N_d$ (see Figures S4
and S5). These two first filters are applied to discard users
having a mobile phone activity that is too low, and/or
not sufficiently staggered over the two-weeks
period. Two time scales have been considered, $N_h$ for the hours
and $N_d$ for the days. The goal is to combine these two
parameters to select users having a significant number of hours
`in activity' spread over a number of days large enough to
guaranty the reliability of the information provided by the most
visited locations extraction process. We were not able to
identify any criteria for calibrating these two parameters, and
we have therefore decided to fix the value of $N_d$ to 4 days
(half of the time period) and the value of $N_h$ to 12 hours.
Combined with a value of $\delta_h=1/3$, this results in keeping
around 50,000 reliable users for each two-weeks period (Figure
\ref{Fig5}a). Later, we will analyze more systematically how the
value of $N_h$ impact the uncertainty in the OD estimation and
its topological structure.

The source code used to extract most visited locations from
individual spatio-temporal trajectories is available online
\footnote{https://github.com/maximelenormand/Most-frequented-locations}.

\begin{figure*}
  \centering 
  \includegraphics[width=12cm]{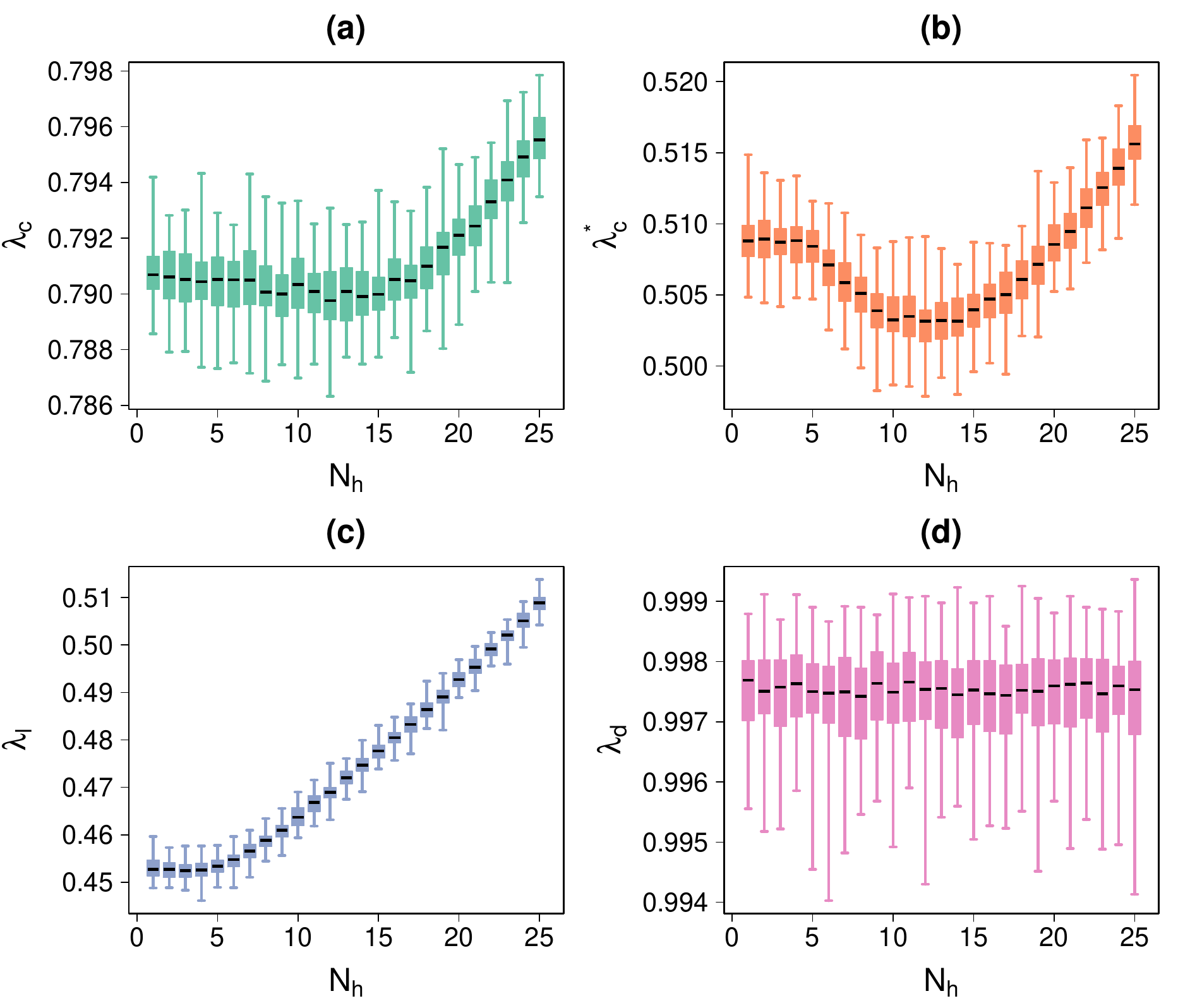}
  \caption{\textbf{Uncertainty when inferring home-work locations
      from mobile phone activity.} Boxplots of the comparisons
    between 100 independent ODs according to $N_h$. (a)
    $\lambda_c$ (all flows). (b) $\lambda_c^*$ (only inter-zonal
    flows). (c) $\lambda_l$. (d) $\lambda_d$. ODs have been
    extracted at the Voronoi scale from independent samples
    composed of 150,000 reliable users' home-work locations (with
    $N_d=4$ and $\delta_h=1/3$). The whiskers correspond to the
    minimum and maximum of the distributions. \label{Fig6}}
\end{figure*}

\subsection*{Sampling strategy and similarity metrics}

\subsubsection*{Sampling strategy}

Similarly to the land use identification, we need a sampling
strategy for assessing the robustness of the OD extracted with
respect to the sample selection, and we use independent samples
of equal sizes drawn at random across the 25 two-weeks
periods. For example, for comparing two ODs retrieved from two
independent samples of 150,000 reliable users' home-work
locations, we draw at random two independent samples of size
6,000 for each of the two-weeks period. Here also, we evaluate
the influence of the sample size and of the spatial resolution on
the uncertainty, by varying the number of signals extracted at
each random extraction and/or by spatially aggregating the
signals over spatial grids of varying sizes.

\begin{figure*}
  \centering 
  \includegraphics[width=13cm]{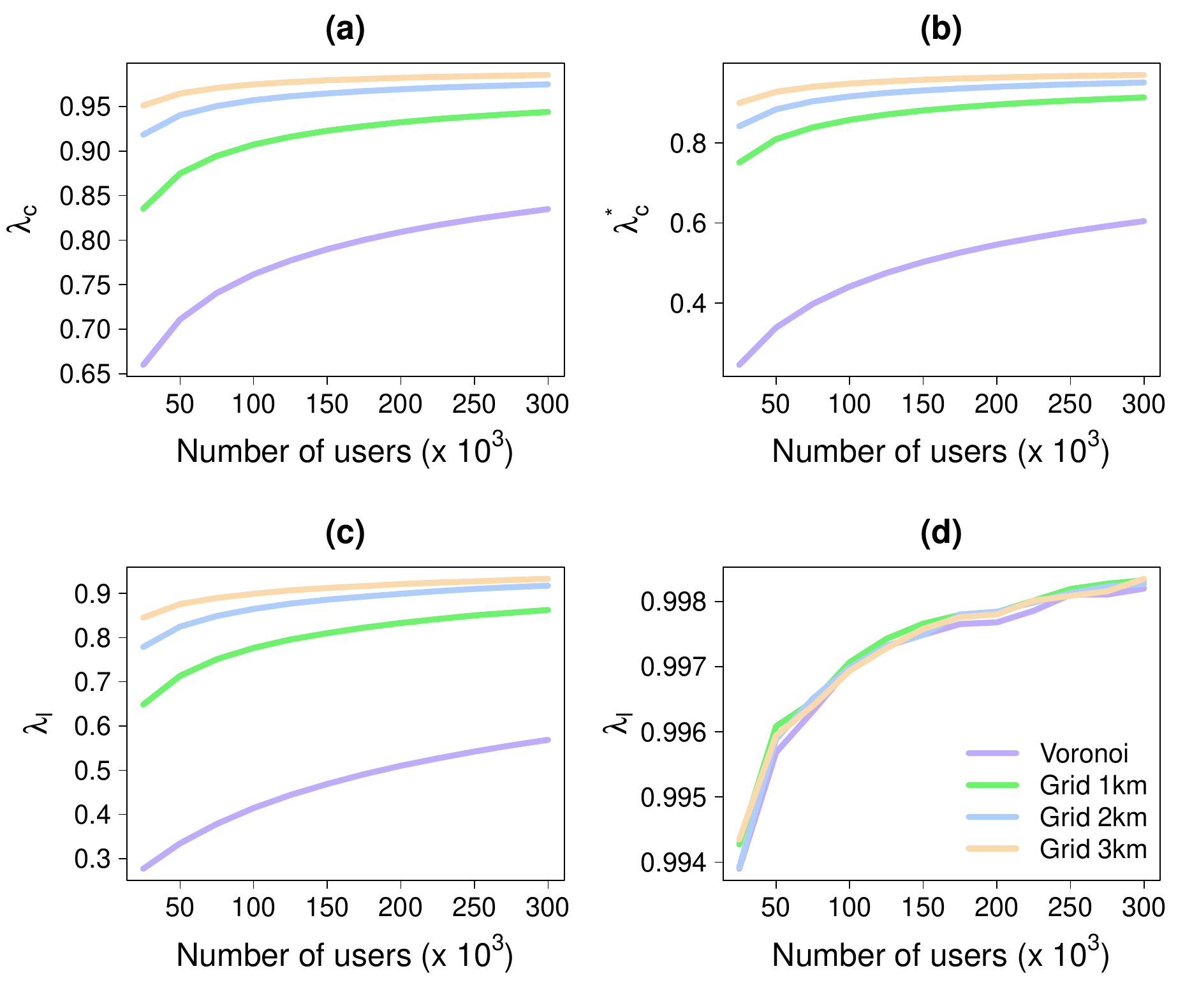}
  \caption{\textbf{Influence of spatial scale and sample size.}
    Similarity metrics as a function of the sample size,
    according to various spatial scales (Voronoi cells and
    regular square grid of cells of 1, 2 and 3 km side
    length). (a) $\lambda_c$ (all flows). (b) $\lambda_c^*$ (only
    inter-zonal flows). (c) $\lambda_l$. (d) $\lambda_d$. The
    values displayed have been averaged over 100 independent
    comparisons with $N_d=4$, $N_h=12$ and $\delta_h=1/3$. The
    same figure displaying the full distribution instead of the
    average is available in Figure S7 in Appendix. \label{Fig7}}
\end{figure*}

\subsubsection*{Similarity metrics}

The resulting commuting networks can be compared using several
similarity metrics, such as the one described in
\cite{Lenormand2016}. We consider two commuting networks $T$ and
$T'$, where $T_{ij}$ is the number of users living in zone $i$
and working in zone $j$, and we will use three different metrics,
that encode different network properties. First, the common
fraction of commuters, noted $\lambda_c$, varying from 0 (when
there is no overlap) to 1 (when the two networks are identical)
is given by
\begin{equation}
  \displaystyle \lambda_c = \frac{2\sum_{i,j} \min (T_{ij},T'_{ij})}{\sum_{i,j} T_{ij} + \sum_{i,j} T'_{ij}} . \label{lambda_c}
 \end{equation}
 With this first metric, the similarity is calculated considering
 all flows, without distinction between the intra-zonal flows and
 the inter-zonal flows. The first type of flows tends to gather a
 large part of the commuters distributed over a limited number of
 links whereas the latter are usually less stable and more
 difficult to estimate. To take into consideration these
 different types of flows we will also consider, as a similarity
 metric, the common fraction of commuters, $\lambda_c^*$, based
 only on the users living and working in two different places
 ($i \neq j$).

 Second, we will consider the common fraction of links,
 $\lambda_l$, that measures similarity in the networks'
 topological structure, and is calculated as
\begin{equation}
  \displaystyle \lambda_l = \frac{2\sum_{i,j} \mathds{1}_{T_{ij}>0} \cdot \mathds{1}_{T'_{ij}>0}}{\sum_{i,j} \mathds{1}_{T_{ij}>0} + \sum_{i,j} \mathds{1}_{T'_{ij}>0}} . \label{lambda_l}
\end{equation}
Third, we measure the common share of commuters according to the
distance, $\lambda_d$, assessing the similarity between commuting
distance distributions and given by
\begin{equation}
  \displaystyle \lambda_d = \frac{\sum_{k} \min (N_{k},N'_{k})}{N} , \label{lambda_d}
\end{equation}
where $N_k$ stands for the number of users with a commuting
distance ranging between $2k-2$ and $2k$ kms ($k$ ranging from 1
to $\infty$) and $N$ for the total number of users.

\subsection*{Uncertainty analysis}

Boxplots of the similarity metric values obtained by comparing
100 independent ODs based on 150,000 reliable users' home-work
locations are displayed in Figure \ref{Fig6}. We first
concentrate on the influence of $N_h$ on the uncertainty and on
the networks' topological structure, and we observe that the only
metrics really affected by the number of hours of activity is the
share of links $\lambda_l$, which increases linearly with
$N_h$. This is an expected behavior since the constraint on the
number of hours in activity selects the most regular individuals
and reduces the noise, decreases the number of links, and
coherently produces more robust ODs. The effect of $N_h$ on the
other similarity metrics is quite negligible, particularly on the
similarity between commuting distance distributions
$\lambda_d$. The values of $\lambda_c$ and $\lambda_c^*$ move up
and down with the number of links (Figure S6 in Appendix) but the
fluctuations stay reasonably small, 0.01 point for the
$\lambda_c$ and 0.02 point for $\lambda_c^*$. The fact that the
number of links increases for values of $N_h$ ranging from 5 to
15 is not trivial (Figure S6 in Appendix). It is even more
surprising to see that this increase is due to the growth of the
number of links of small sizes (between 2 and 10 commuters) that
are not able to counterbalance the falls of the number of very
small (1 user) and big links (more than 10 users). These changes
of network structure are not easy to understand and would
probably necessitate more information about commuting in Senegal
to be investigated more thoroughly.

Regarding the uncertainty, the results are not completely
conclusive, the commuting networks show a good agreement with
around 79\% of commuters in common (considering both inter- and
intra-zonal flows), but with a value that falls down to 50\% when
only inter-zonal flows are considered. This decrease of
similarity can be explain by the fact that intra-zonal flows
represent 60\% of the commuters distributed on a number of links
that is bounded by the number of zones, whereas inter-zonal flows
are generally smaller since they represent less commuters
distributed on a higher number of links. This is why when we
remove the intra-zonal flows which are usually easier to
estimate, we increase the uncertainty.  The $\lambda_l$ values
are also quite low, between 45\% and 51\% of links are in common
between the different networks according to the value of
$N_h$. An encouraging result is that the common part of commuters
according to the distance is very high, showing around 99\% of
similarity between the commuting distance distributions. However,
it is important to keep in mind that these mixed results are
obtained with a few thousand users for each two-weeks commuting
network, drawn at a high spatial resolution with an average
surface area equal to 0.5 km$^2$.

\subsection*{Influence of scale and sample size}

The effect of the spatial resolution and sample sizes on the
similarity metrics can be investigated by varying the number of
reliable users' home-work most visited locations to build the ODs
and/or by aggregating them spatially using grid cells of
different sizes (see \cite{Lenormand2014} for more details about
the aggregation method based on the area of the intersection
between the Voronoi and the grid cells). As it can be observed
in Figure \ref{Fig7}, increasing the sample size and/or the scale
greatly improve the results. Here again, considering at least
100,000 reliable users seems to be a good trade-off and ensure a
common part of commuters larger than 75\%. The most significant
improvement comes from the spatial aggregation, which at least
double $\lambda_c^*$ and $\lambda_l$ values and allow us to
obtain $\lambda_c$ values almost always larger than 0.85. There
is one exception, though, with $\lambda_l$ that seems to be
independent from the spatial aggregation scale used.

\subsection*{Temporal variations}

\begin{figure}[!h]
  \centering 
  \includegraphics[width=\linewidth]{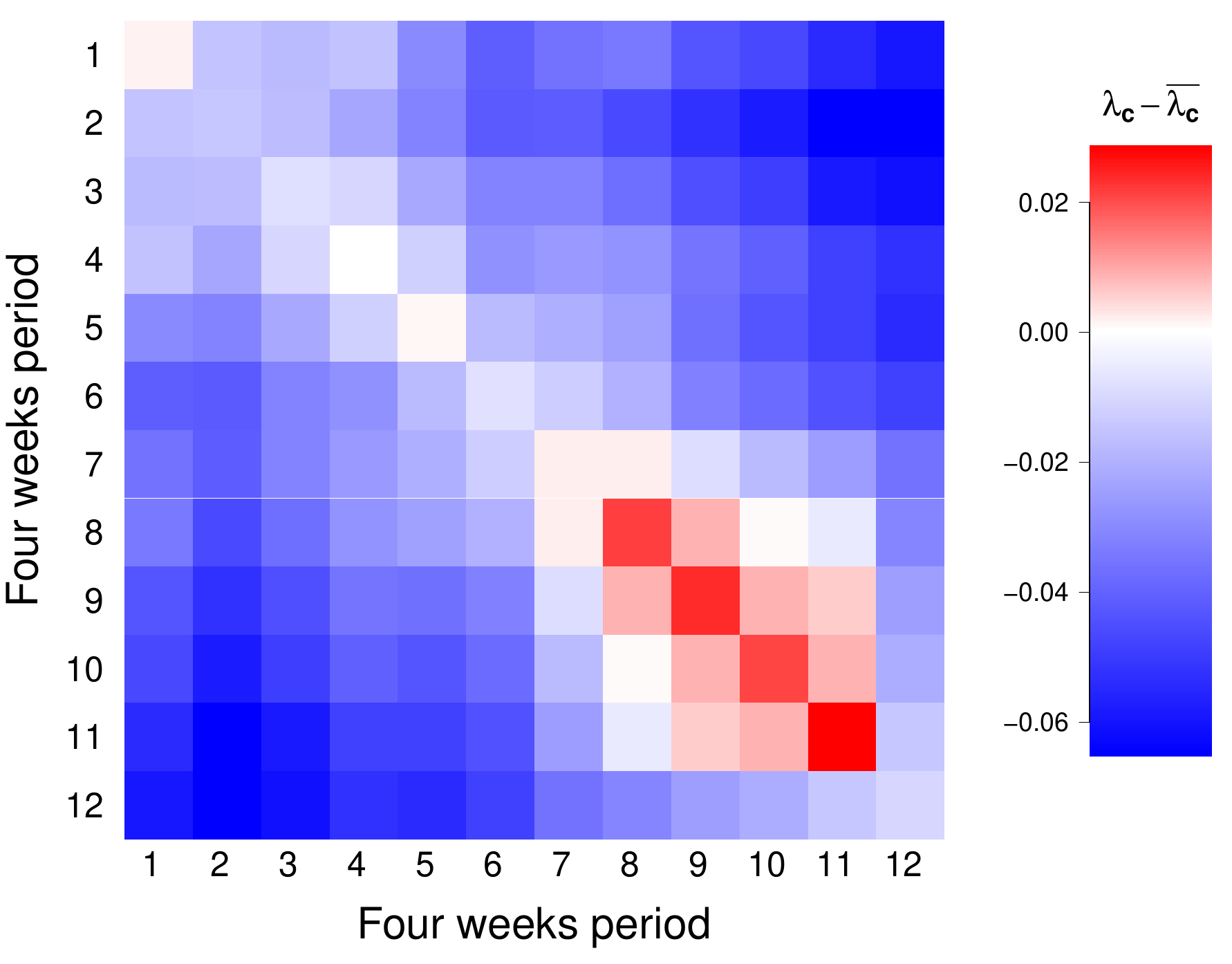}
  \caption{\textbf{Temporal variations.} Standard $\lambda_c$
    between ODs extracted for different time windows (four weeks
    periods, from the first week of January to the last week of
    November). The standardization has been obtained by
    subtracting the average $\lambda_c$ obtained by comparing 100
    independent samples based on 50,000 reliable users' home-work
    most visited locations drawn at random across the entire
    year. The results have been averaged over 100 independent
    comparisons based on 50,000 reliable users' home-work most
    visited locations at the Voronoi scale. Plots of the other
    similarity metrics are available in Figure S8 in
    Appendix. \label{Fig8}}
\end{figure}

The effect of temporal variations on the uncertainty of OD
estimates can be investigated by considering 12 consecutive time
windows of four weeks (from the first week of January to the last
week of November, see section on land use detection). In this
case, ODs are based on 50,000 reliable users' home-work most
visited locations, determined at the Voronoi scale. The
similarity metrics have been standardized by subtracting the
average metrics values across the entire year. Figure \ref{Fig8}
shows the results obtained with the common part of commuters
$\lambda_c$ (similar results obtained with the other metrics are
shown on Figure S8 in Appendix). As in the case of land use
identification, the values are close to the average, lower for
the inter time windows comparisons than for the intra
comparisons. We also observe that the results are more stable
after summer holidays, during the second half of the year. An
evolution of the commuting networks over time can be observed
with a decrease of the similarity metrics values as the time
elapsed between time periods increases.
 
\subsection*{Sampling of points along individual trajectories}

\begin{figure}[!h]
  \centering 
  \includegraphics[width=\linewidth]{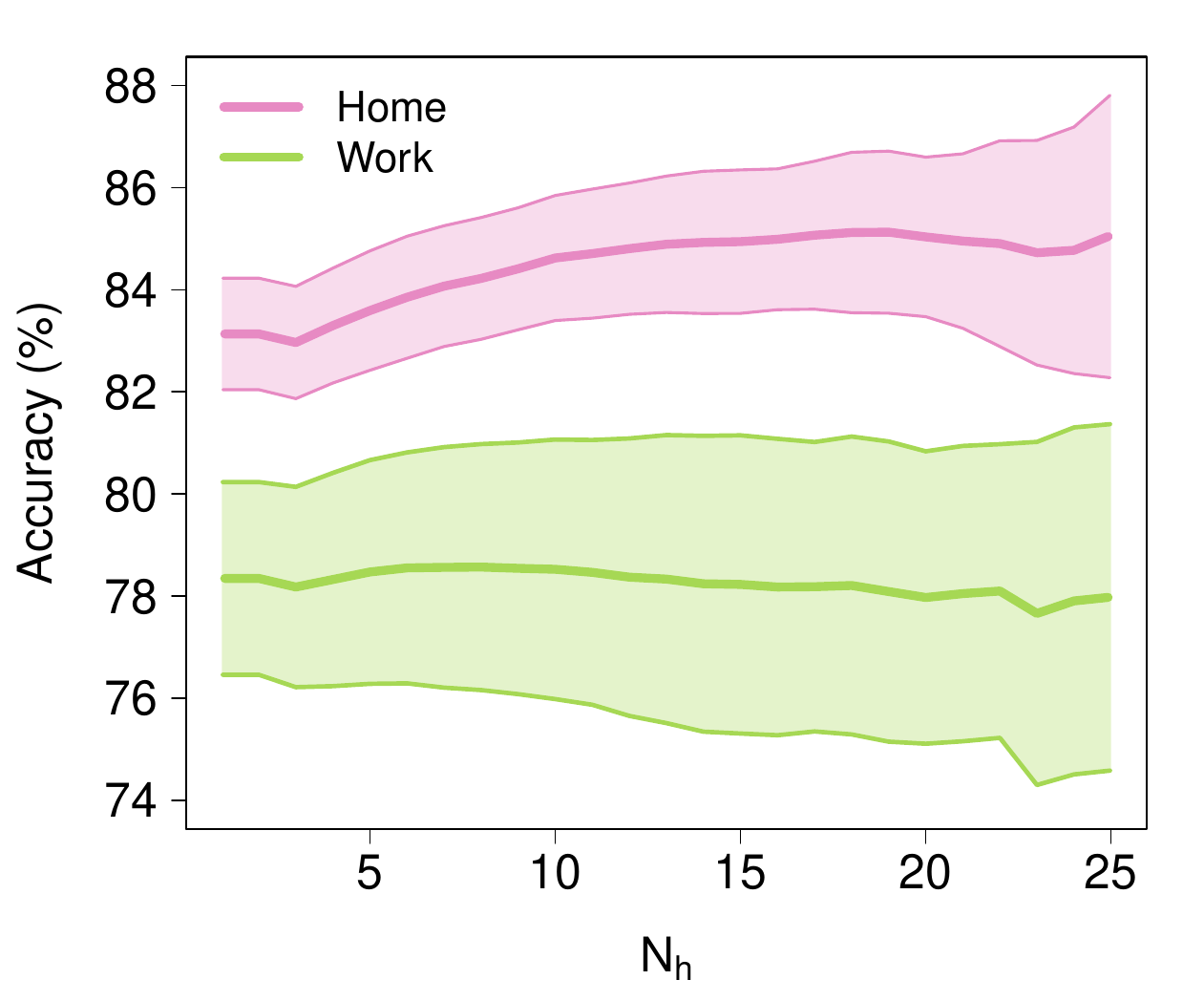}
  \caption{\textbf{Percentage of accuracy in home and work
      locations' estimation according to $N_h$.} Results have
    been averaged over the 25 two-weeks time windows. The thin
    lines represent one standard deviation. \label{Fig9}}
\end{figure}

In order to go further, for each of the 25 two-weeks periods and
for each user, we identified the home and work locations for each
of the two weeks considered independently, by following the
procedure described above with $N_d=2$ (half of the time window)
and $\delta_h=1/3$. This allowed us to assess the influence of
the sampling of points along individual trajectories when
identifying the home and work locations. We then compared the
locations identified for each of the two-weeks period according
to the value of $N_h$. While the effect of $N_h$ on the level of
accuracy is quite low, it is however important to note that the
level of accuracy in the estimate of home location tends to
increase with $N_h$. Considering the high spatial resolution and
the small time window, a good agreement is obtained, with an
accuracy of around 84\% for home (average over the 25 two-week
periods) and 78\% for work (Figure \ref{Fig9}). Moreover, 50\% of
the inaccurate locations are less than 2 kms distant from each
other. We can therefore conclude that the identification of
users' home and work locations from mobile phone activity also
shows a high level of robustness to sample selection.

\section*{Discussion}

Data passively produced through information and communications
technologies have been increasingly used by researchers since the
middle of the 2000's % for analyzing a variety of human processes
% and activities. In particular, 
and our understanding of number of aspects of human mobility has
already been deeply renewed thanks to these new data
sources. More generally, the longitudinal tracking of anonymized
individuals opens the door to an enhanced understanding of social
phenomena that could not be studied empirically at such scales
before. However, these data obviously suffer from a number of
biases \cite{Lewis2015}, which include in particular sample
selection. Systematic tests are then required to characterize
statistical validity, along with cross-checks between various
data sources.

This in mind, we performed two uncertainty analysis of results
obtained with mobile phone data produced by millions of
anonymized individuals during an entire year. In the first part
of the analysis, we assessed the uncertainty when inferring
land use from human activity, estimated from the
number of mobile phone users present in different parts of the
city, at different moments of the week. A good agreement was
obtained between land uses identified from independent and
randomly selected samples of individuals mobile phone activity
 during one week. We showed that samples should be composed of at least 100,000
individual signals of activity in order to ensure that there is
at least 75\% of shared surface area between the spatial
partitions of land uses automatically retrieved, whatever the
level of spatial aggregation of the data. In the second part of
the analysis, we investigated the influence of sample selection
on the identification of users' home and work locations. We first
examined the impact of the selection of users on the
journey-to-work commuting networks extracted at the city
scale. In our case-study of the city of Dakar, we showed that the
level of uncertainty is largely dependent on the spatio-temporal
resolution, and that good results were reachable with a
reasonable level of spatial aggregation. We then analyzed the
effect of the sampling of locations along mobile phone users'
trajectories on the identification of their home and work
locations. Most of the locations identified with different
samples were the same, or very close to one another. Finally, we
also showed that in both cases the uncertainty varies according
to the period of the year at which the data were gathered,
particularly in summer for the land use identification.

A useful and natural extension to this work would be to compare
the results obtained in this study with traditional mobility data
sources such as surveys or census data, particularly to calibrate
the parameters of the home-work most visited location
identification process. It was unfortunately not possible to
obtain such data from the Senegal census, but this type of
analysis should be done whenever possible.

For these two spatial information retrieval tasks, our results
suggest that the level of uncertainty associated with sample
selection is globally low. Further work in this direction include
the reproduction of such uncertainty analysis with other datasets
coming from different countries and data sources. An important
aspect of the rapidly growing `new science of cities'
\cite{Batty2013}, which heavily relies on new data sources, is to
be able to reproduce results with different datasets, and to
characterize and control to what extent the information provided
by different sources are biased in a particular direction.

More studies in this spirit need to be done to strengthen the
foundations of the field dedicated to the understanding of urban
mobility and urban dynamics through ICT data. From a publication
point of view, trying to reproduce previous results with
different data sources, or to estimate the robustness of
previously published results, might not be as appealing as
proposing new measures and models, but is crucially important as
well. 

\section*{Acknowledgments}

Partial financial support has been received from the Spanish
Ministry of Economy (MINECO) and FEDER (EU) under project
ESOTECOS (FIS2015-63628-C2-2-R), and from the EU Commission
through project INSIGHT. The work of TL has been funded under the
PD/004/2015, from the Conselleria de Educaci\'on, Cultura y
Universidades of the Government of the Balearic Islands and from
the European Social Fund through the Balearic Islands ESF
operational program for 2013-2017.

\bibliographystyle{unsrt}
\bibliography{ictdataccuracy}

\begin{thebibliography}{10}

\bibitem{Kaisler2013}
S.~Kaisler, F.~Armour, J.~A. Espinosa, and W.~Money.
\newblock Big {Data}: {Issues} and {Challenges} {Moving} {Forward}.
\newblock In {\em 2014 47th {Hawaii} {International} {Conference} on {System}
  {Sciences}}, volume~0, pages 995--1004, 2013.

\bibitem{Lewis2015}
K.~Lewis.
\newblock Three fallacies of digital footprints.
\newblock {\em Big Data \& Society}, 2(2), 2015.

\bibitem{Ratti2006}
C.~Ratti, D.~Frenchman, Riccardo~M. Pulselli, and S.~Williams.
\newblock Mobile landscapes: using location data from cell phones for urban
  analysis.
\newblock {\em Environment and Planning B: Planning and Design},
  33(5):727--748, 2006.

\bibitem{Louail2014}
T.~Louail, M.~Lenormand, O.~G. Cantu, M.~Picornell, R.~Herranz,
  E.~Fr\'{\i}as-Mart\'{\i}ne, J.~J. Ramasco, and M.~Barthelemy.
\newblock From mobile phone data to the spatial structure of cities.
\newblock {\em Scientific reports}, 4, 2014.

\bibitem{Calabrese2015}
F.~Calabrese, L.~Ferrari, and V.~D Blondel.
\newblock Urban sensing using mobile phone network data: a survey of research.
\newblock {\em ACM Computing Surveys (CSUR)}, 47(2):25, 2015.

\bibitem{Louail2015}
T.~Louail, M.~Lenormand, M.~Picornell, O.~G. Cant{\'u}, R.~Herranz,
  E.~Fr\'{\i}as-Mart\'{\i}ne, J.~J. Ramasco, and M.~Barthelemy.
\newblock Uncovering the spatial structure of mobility networks.
\newblock {\em Nature Communications}, 6, 2015.

\bibitem{Lenormand2014}
M.~Lenormand, M.~Picornell, O.~G. Cant{\'u}-Ros, A.~Tugores, T.~Louail,
  R.~Herranz, M.~Barthelemy, E.~Fr{\'i}as-Mart{\'i}nez, and J.~J. Ramasco.
\newblock Cross-{Checking} {Different} {Sources} of {Mobility} {Information}.
\newblock {\em PLoS ONE}, 9(8):e105184, 2014.

\bibitem{Schneider2013}
C.~M. Schneider, V.~Belik, T.~Couronn{\'e}, Z.~Smoreda, and M.~C. Gonz{\'a}lez.
\newblock Unravelling daily human mobility motifs.
\newblock {\em Journal of The Royal Society Interface}, 10(84):20130246, 2013.

\bibitem{Tizzoni2014}
M.~Tizzoni, P.~Bajardi, A.~Decuyper, G.~K.~K. King, C.~M. Schneider,
  V.~Blondel, Z.~Smoreda, M.~C. Gonz{\'a}lez, and V.~Colizza.
\newblock On the {{Use}} of {{Human Mobility Proxies}} for {{Modeling
  Epidemics}}.
\newblock {\em PLOS Comput Biol}, 10(7):e1003716, 2014.

\bibitem{Deville2014}
P.~Deville, C.~Linard, S.~Martin, M.~Gilbert, F.~R. Stevens, A.~E. Gaughan,
  V.~D. Blondel, and A.~J. Tatem.
\newblock Dynamic population mapping using mobile phone data.
\newblock {\em Proceedings of the National Academy of Sciences},
  111(45):15888--15893, 2014.

\bibitem{Alexander2015}
L.~Alexander, S.~Jiang, M.~Murga, and M.~C. Gonz{\'a}lez.
\newblock Origin{\textendash}destination trips by purpose and time of day
  inferred from mobile phone data.
\newblock {\em Transportation Research Part C: Emerging Technologies}, 58, Part
  B:240--250, 2015.

\bibitem{Toole2015}
J.~L. Toole, S.~Colak, B.~Sturt, L.~P. Alexander, A.~Evsukoff, and M.~C.
  Gonz{\'a}lez.
\newblock The path most traveled: {Travel} demand estimation using big data
  resources.
\newblock {\em Transportation Research Part C: Emerging Technologies}, 58, Part
  B:162--177, 2015.

\bibitem{Jiang2016}
S.~Jiang, Y.~Yang, S.~Gupta, D.~Veneziano, S.~Athavale, and M.~C. Gonz{\'a}lez.
\newblock The {TimeGeo} modeling framework for urban motility without travel
  surveys.
\newblock {\em Proceedings of the National Academy of Sciences}, page
  201524261, 2016.

\bibitem{Montjoye2014}
Y.-A. de~Montjoye, Z.~Smoreda, R.~Trinquart, C.~Ziemlicki, and V.~D. Blondel.
\newblock {D4D-Senegal: The Second Mobile Phone Data for Development
  Challenge}.
\newblock {\em arXiv preprint}, arXiv:1407.4885, 2014.

\bibitem{Soto2011}
V.~Soto and E.~Fr\'{\i}as-Mart\'{\i}nez.
\newblock Automated land use identification using cell-phone records.
\newblock In {\em Proceedings of the 3rd ACM international workshop on
  MobiArch}, HotPlanet '11, pages 17--22, New York, NY, USA, 2011. ACM.

\bibitem{Frias2012}
V.~Fr\'{\i}as-Mart\'{\i}nez, V.~Soto, H.~Hohwald, and
  E.~Fr\'{\i}as-Mart\'{\i}nez.
\newblock Characterizing urban landscapes using geolocated tweets.
\newblock In {\em SocialCom/PASSAT}, pages 239--248. IEEE, 2012.

\bibitem{Toole2012}
J.~L. Toole, M.~Ulm, M.~C. Gonz{\'a}lez, and D.~Bauer.
\newblock Inferring {{Land Use}} from {{Mobile Phone Activity}}.
\newblock In {\em Proceedings of the {{ACM SIGKDD International Workshop}} on
  {{Urban Computing}}}, UrbComp '12, pages 1--8, New York, NY, USA, 2012.
  {ACM}.

\bibitem{Pei2013}
T.~Pei, S.~Sobolevsky, C.~Ratti, S.~L. Shaw, and C.~Zhou.
\newblock A new insight into land use classification based on aggregated mobile
  phone data.
\newblock {\em International Journal of Geographical Information Science},
  28:1988--2007, 2014.

\bibitem{Lenormand2015}
M.~Lenormand, M.~Picornell, O.~Garcia~Cant{\'u}, A.~Tugores, T.~Louail,
  R.~Herranz, M.~Barthelemy, E.~Fr{\'i}as-Mart{\'i}nez, and J.~J. Ramasco.
\newblock Comparing and modeling land use organization in cities.
\newblock {\em Royal Society Open Science}, 2:150459, 2015.

\bibitem{Ahas2010}
R.~Ahas, Olle Silm, S.and~J., E.~Saluveer, and M.~Tiru.
\newblock Using {Mobile} {Positioning} {Data} to {Model} {Locations}
  {Meaningful} to {Users} of {Mobile} {Phones}.
\newblock {\em Journal of Urban Technology}, 17(1):3--27, 2010.

\bibitem{Isaacman2011}
S.~Isaacman, R.~Becker, R.~C{\'a}ceres, S.~Kobourov, M.t Martonosi, J.~Rowland,
  and A.~Varshavsky.
\newblock Identifying {Important} {Places} in {People}'s {Lives} from
  {Cellular} {Network} {Data}.
\newblock In Kent Lyons, Jeffrey Hightower, and Elaine~M. Huang, editors, {\em
  Pervasive {Computing}}. Springer Berlin Heidelberg, 2011.

\bibitem{Rosvall2008}
M.~Rosvall and C.~T. Bergstrom.
\newblock Maps of random walks on complex networks reveal community structure.
\newblock {\em Proceedings of the National Academy of Sciences},
  105(4):1118--1123, 2008.

\bibitem{Note1}
https://github.com/maximelenormand/Most-frequented-locations.

\bibitem{Lenormand2016}
M.~Lenormand, A.~Bassolas, and J.~J. Ramasco.
\newblock Systematic comparison of trip distribution laws and models.
\newblock {\em Journal of Transport Geography}, 51:158--169, 2016.

\bibitem{Batty2013}
M.~Batty.
\newblock {\em The New Science of Cities}.
\newblock MIT Press, 2013.

\end{thebibliography}

\onecolumngrid
\vspace*{2cm}
\newpage
\onecolumngrid

\makeatletter
\renewcommand{\fnum@figure}{\sf\textbf{\figurename~\textbf{S}\textbf{\thefigure}}}
\renewcommand{\fnum@table}{\sf\textbf{\tablename~\textbf{S}\textbf{\thetable}}}
\makeatother

\setcounter{figure}{0}
\setcounter{table}{0}
\setcounter{equation}{0}

\section*{Appendix}

\vspace{5cm}
\begin{figure}[!ht]
  \centering 
  \includegraphics[width=\linewidth]{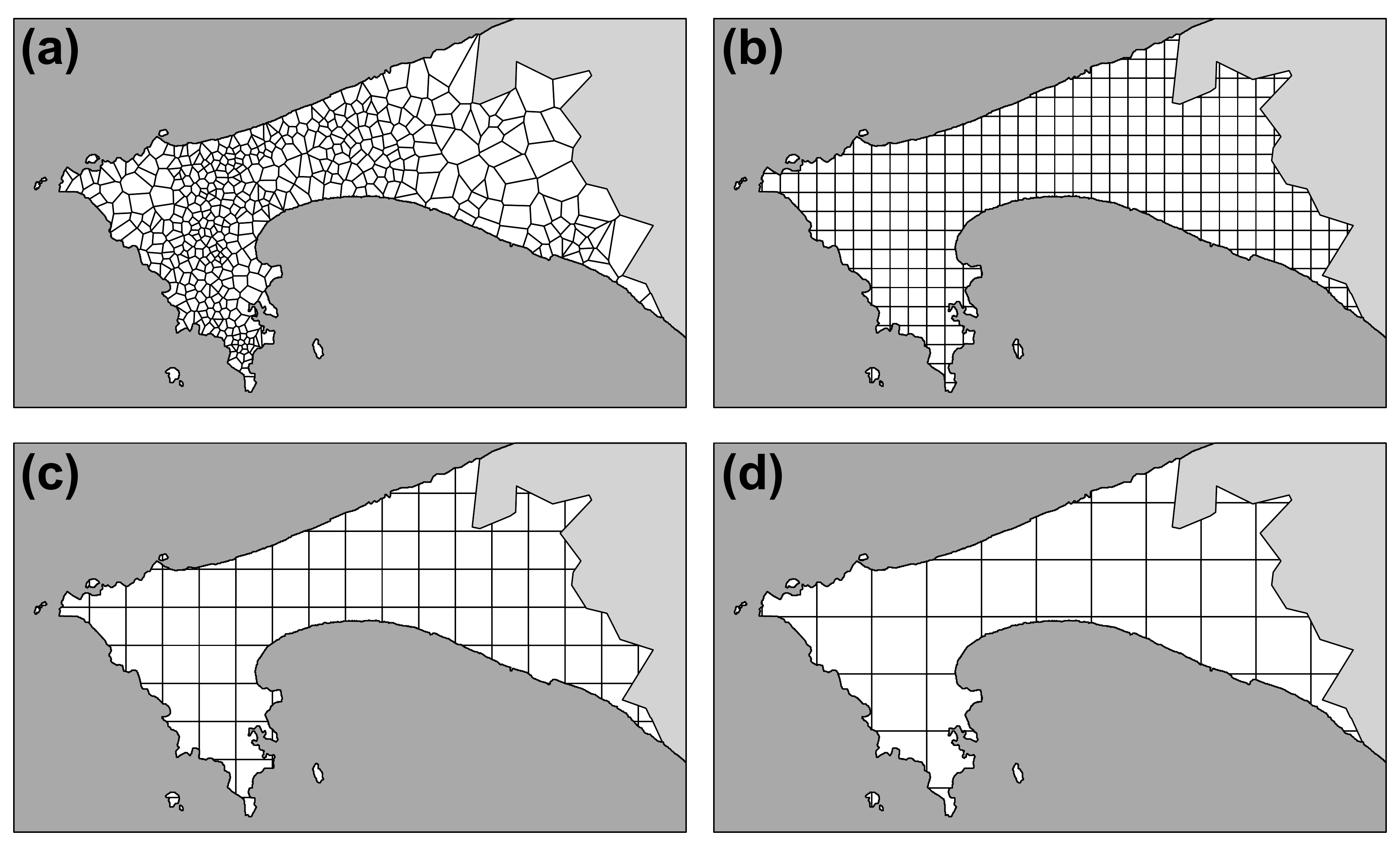}
  \caption{\textbf{Map of the Dakar region.} The white area represents the region of interest and the dark gray area represents the sea. Three scales of geographic units are represented. (a) Voronoi cells. (b) Grid cells ($1\times1 \mbox{ km}^2$). (c) Grid cells ($2\times2 \mbox{ km}^2$). (d) Grid cells ($3\times3 \mbox{ km}^2$). \label{FigS1}}
\end{figure}

\begin{figure}
  \centering 
  \includegraphics[width=15cm]{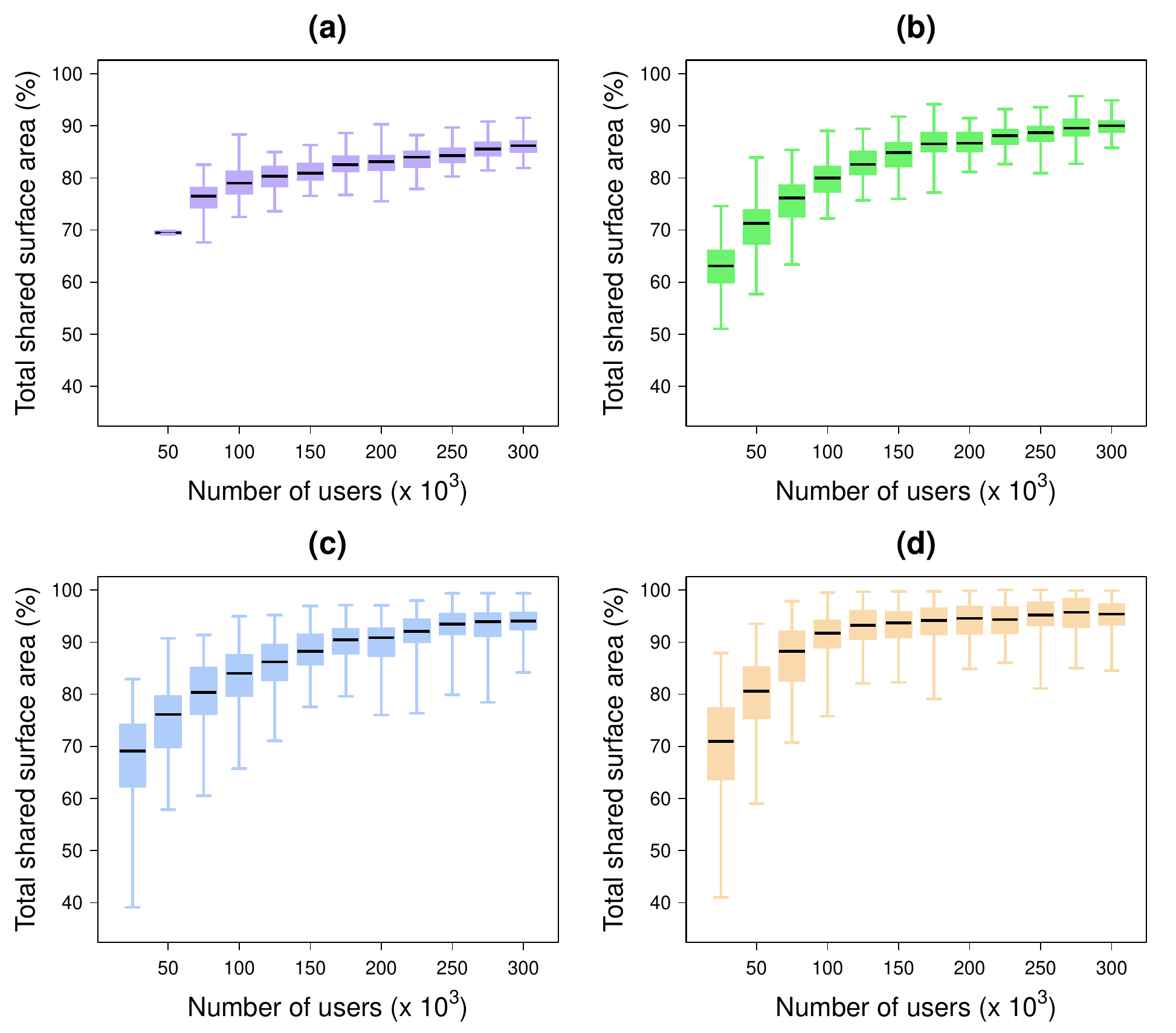}
  \caption{\textbf{Influence of scale and sample size.} Boxplots of the total share surface area as a function of the sample size according to the spatial scale. (a) Voronoi cells. (b) Grid cells ($1\times1 \mbox{ km}^2$). (c) Grid cells ($2\times2 \mbox{ km}^2$). (d) Grid cells ($3\times3 \mbox{ km}^2$). These results are based on 100 independent comparisons. The whiskers correspond to the minimum and maximum of the distributions. \label{FigS2}}
\end{figure}

\begin{figure}
  \centering 
  \includegraphics[width=15cm]{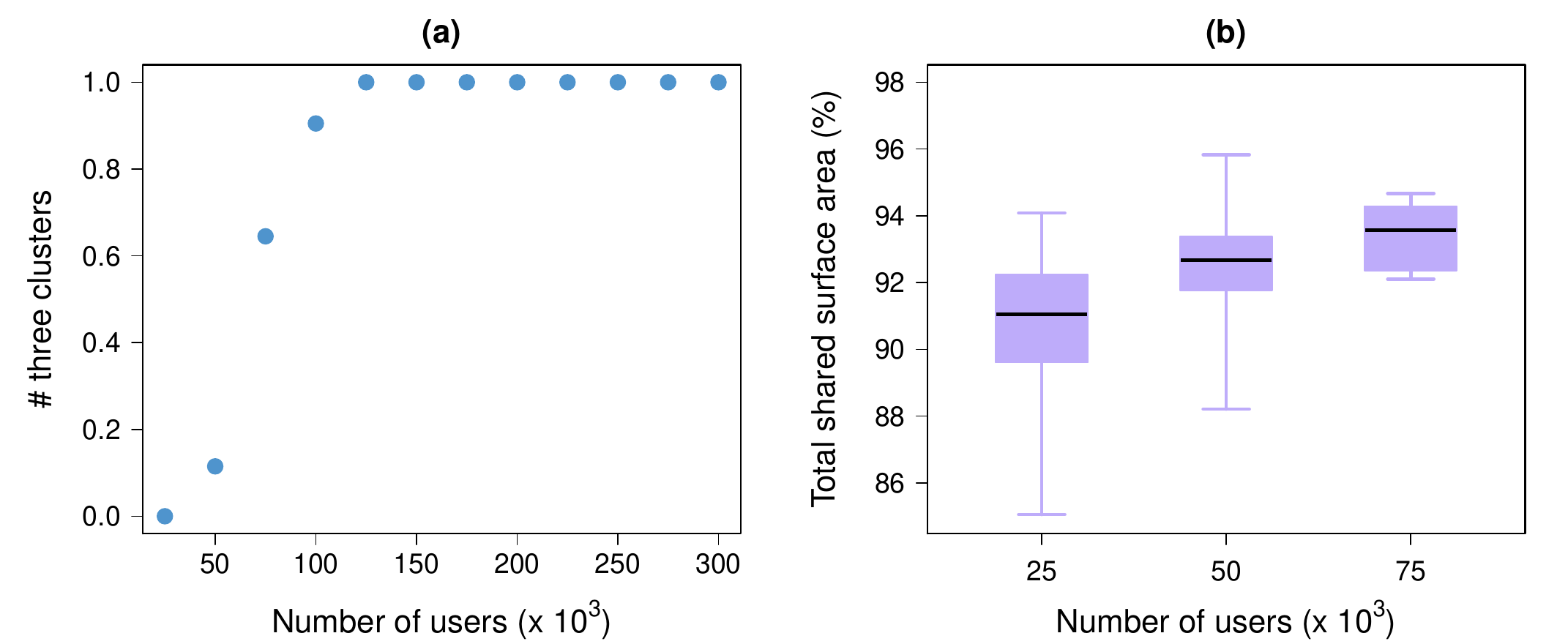}
  \caption{\textbf{Influence of the sample size (at the Voronoi scale) on the number of clusters.} (a) Fraction of random extractions for which three clusters were detected. (b) Boxplots of the share surface area for comparisons for which the number of clusters detected in the two independent samples were two. The whiskers correspond to the minimum and maximum of the distributions.\label{FigS3}}
\end{figure}

\begin{figure}
  \centering 
  \includegraphics[width=15cm]{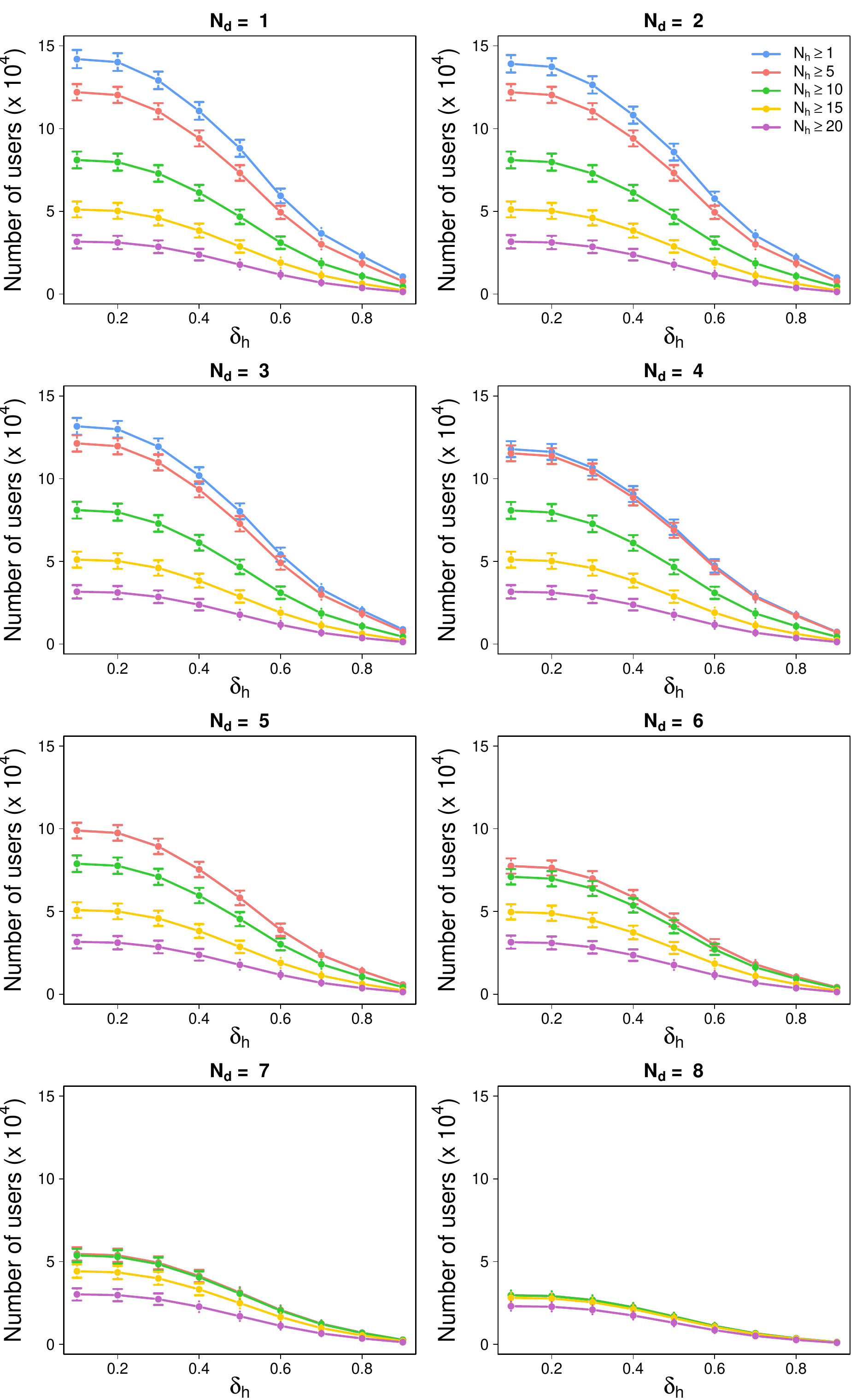}
  \caption{\textbf{Influence of the parameters on the number of reliable users.} Number of reliable users as a function of $\delta_h$ and according to $N_h$ and $N_d$. Only people living and working in the region of Dakar have been considered. The values have been averaged over the 25 weeks, and the error bars represent the standard deviation.\label{FigS4}}
\end{figure}

\begin{figure}
  \centering 
  \includegraphics[width=15cm]{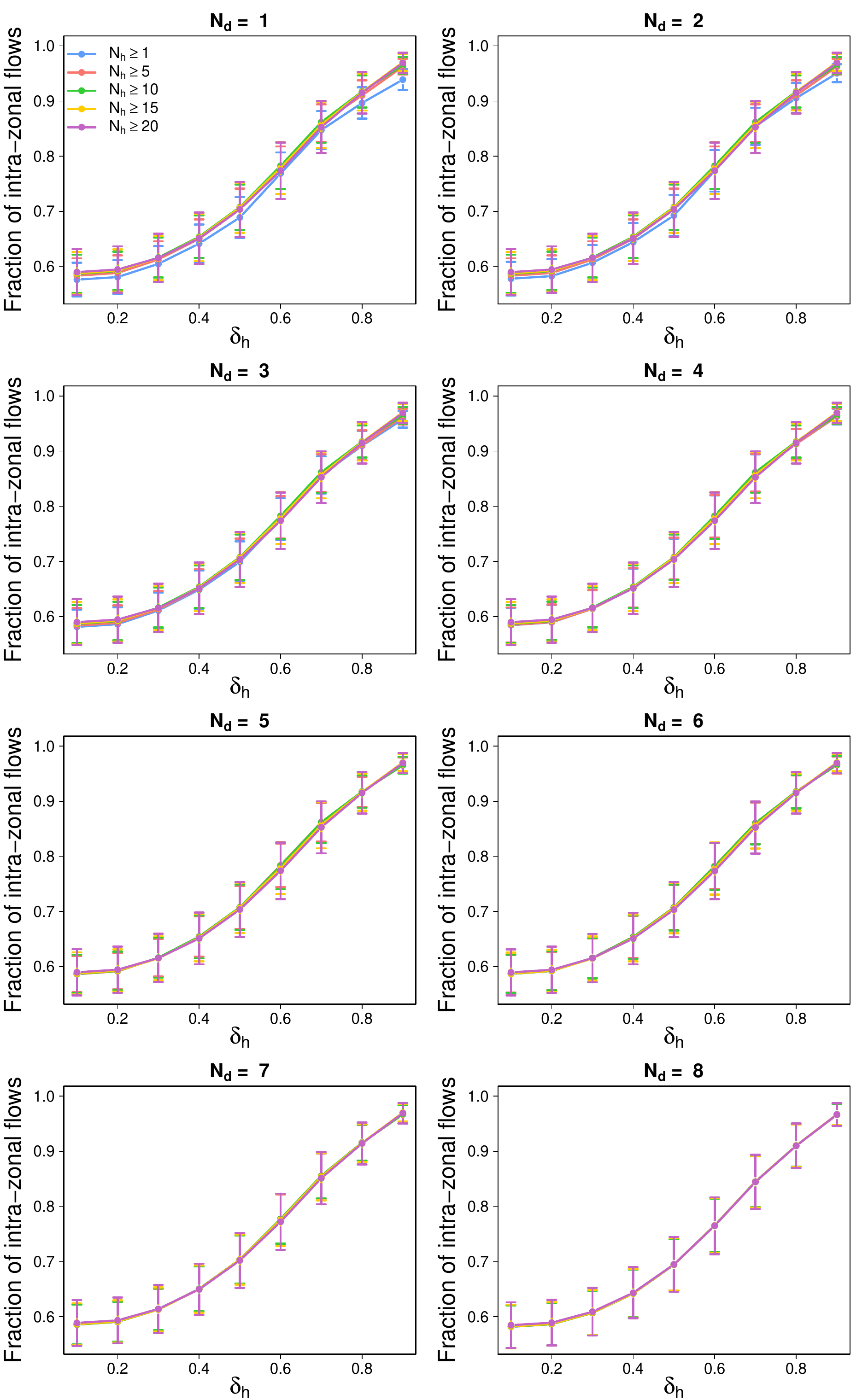}
  \caption{\textbf{Influence of the parameters on the fraction of intra-zonal flows.} Fraction of users living and working in the same zone as a function of $\delta_h$ and according to $N_h$ and $N_d$. Only people living and working in the region of Dakar have been considered. The values have been averaged over the 25 weeks, and the error bars represent the standard deviation.\label{FigS5}}
\end{figure}

\begin{figure}
  \centering 
  \includegraphics[width=13cm]{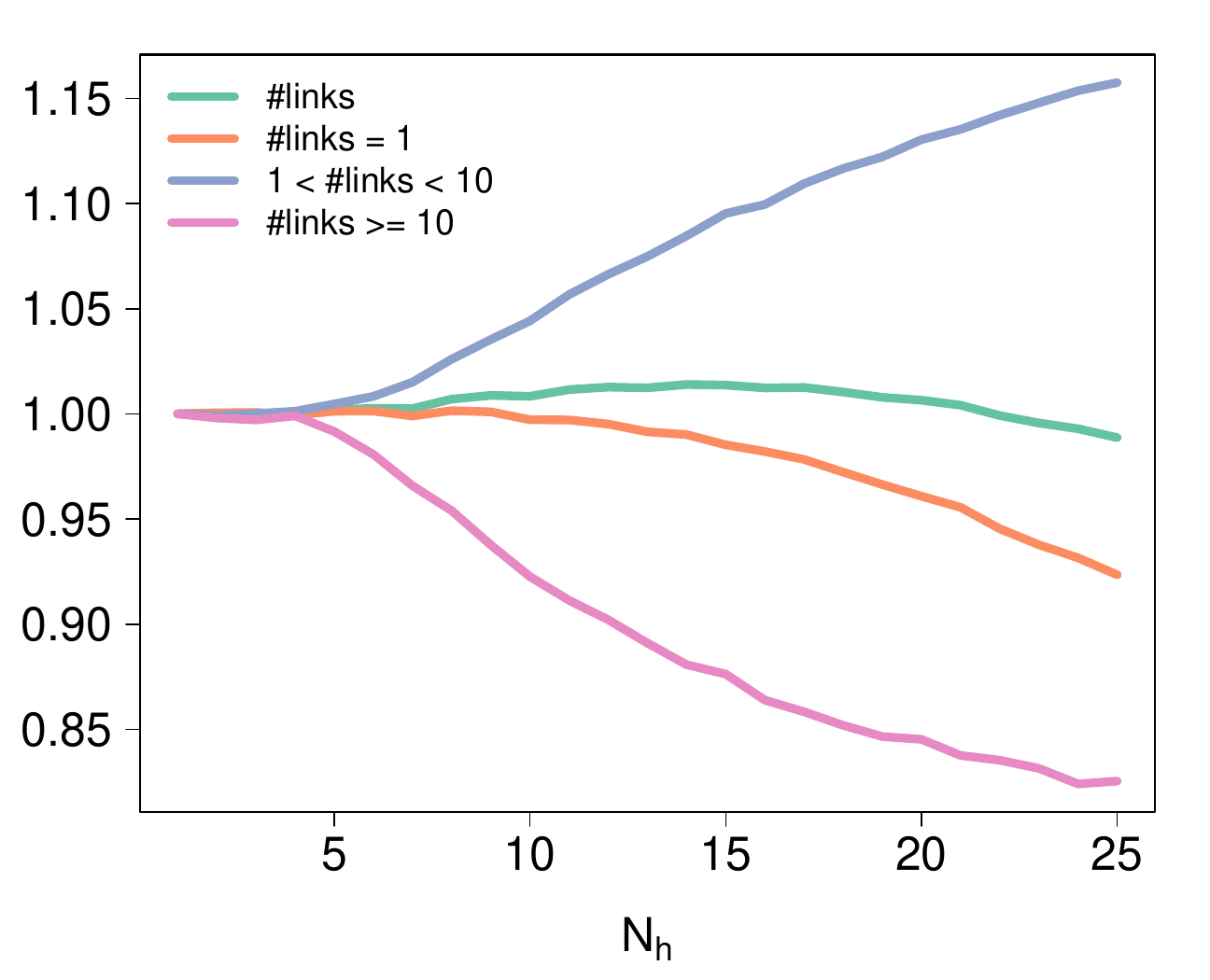}
  \caption{\textbf{Influence of $N_h$ on the number of links.} The value have been normalized by the value obtained with $N_h=1$. \label{FigS6}}
\end{figure}

\begin{figure}
  \centering 
  \includegraphics[width=13cm]{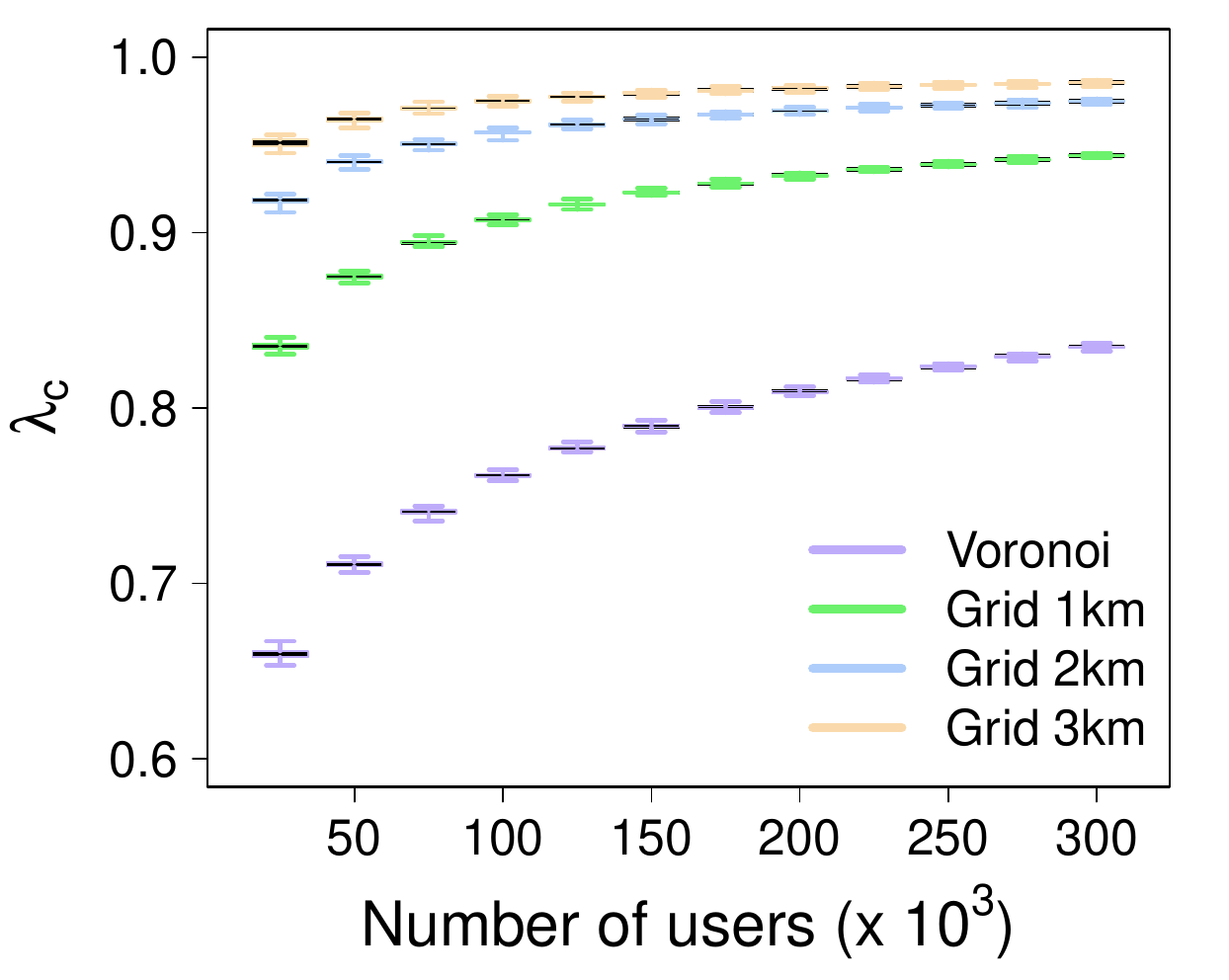}
  \caption{\textbf{Influence of scale and sample size.} Boxplots of $\lambda_c$ as a function of the sample size according to the spatial scale (Voronoi cells and grid cells of 1, 2 and 3 km side length). The distributions are based on 100 independent comparisons with $N_d=4$, $N_h=12$ and $\delta_h=1/3$. The whiskers correspond to the minimum and maximum of the distributions. \label{FigS7}}
\end{figure}

\begin{figure}[!h]
  \centering 
  \includegraphics[width=\linewidth]{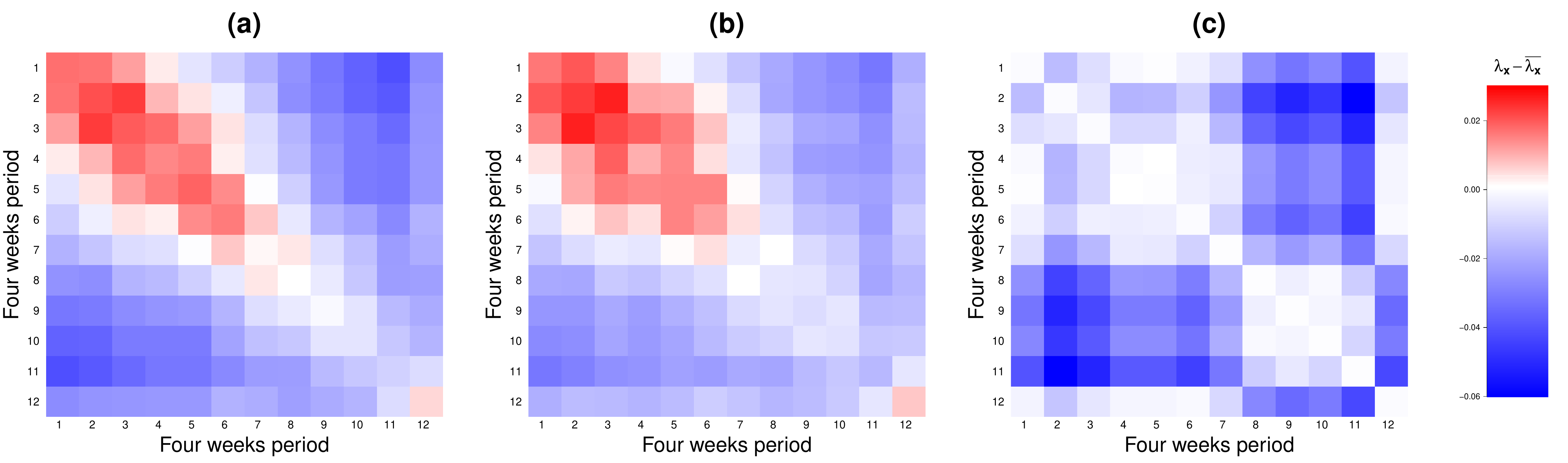}
  \caption{\textbf{Temporal variations variations.} Standard similarity metrics between ODs extracted at different time windows (four weeks periods, from the first week of January to the last week of November). (a) $\lambda_c^*$. (b) $\lambda_l$. (c) $\lambda_d$. The standardization has been obtained by subtracting average values obtained by comparing 100 independent samples based on 50,000 reliable users' home-work most visited locations drawn at random across the entire year. The values have been averaged over 100 independent comparisons based on 50,000 reliable users' home-work most visited locations at the Voronoi scale. \label{FigS8}}
\end{figure}
 
\end{document}